\documentclass[aps]{revtex4-2}
\usepackage{graphicx}

\usepackage[utf8]{inputenc}
\usepackage[T1]{fontenc}
\usepackage{bm, amsmath,amssymb,color}
\usepackage{caption}
\usepackage{subcaption}
\usepackage{svg}
\usepackage{soul}
\usepackage{soul}

\newcommand{\bnabla}{\boldsymbol{\nabla}}
\newcommand{\bcdot}{\boldsymbol{\cdot}}
\newcommand{\lan}{\left\langle}
\newcommand{\ran}{\right\rangle}

\usepackage{xcolor}
\begin{document}

\preprint{APS/123-QED}

\title{Impact of pressure anisotropy on the cascade rate of Hall-MHD turbulence with biadiabatic ions}

\author{Pauline A. Simon}
\email{pauline.simon@qmul.ac.uk}
\affiliation{Department of Physics and Astronomy, Queen Mary University of London, London E1 4NS, United Kingdom}

\author{Fouad Sahraoui, Sébastien Galtier}
\affiliation{Laboratoire de Physique des Plasmas (LPP), CNRS, Observatoire de Paris, Sorbonne Université, Université Paris-Saclay, École Polytechnique, Institut Polytechnique de Paris, 91120 Palaiseau, France}

\author{Dimitri Laveder, Thierry Passot, Pierre-Louis Sulem}
\affiliation{Université Côte d’Azur, Observatoire de la Côte d’Azur, CNRS, Laboratoire J.L. Lagrange, Boulevard de l’Observatoire, CS 34229, F-06304 Nice Cedex 4, France}

 \date{\today}

\begin{abstract} 
The impact of ion pressure anisotropy on the energy cascade rate of Hall-MHD turbulence with biadiabatic ions and isothermal electrons is evaluated in three-dimensional direct numerical simulations, using the exact (or third-order) law derived in \citet{simon_exact_2022}.
It is shown that pressure anisotropy can enhance or reduce the cascade rate, depending on the scales, in comparison with the prediction of the exact law with isotropic pressure, by an amount that correlates well with pressure anisotropy $a_p=\frac{p_\perp}{p_\parallel}\neq1$ that develops in simulations initialized with an isotropic pressure (${a_p}_0=1$). A simulation with initial pressure anisotropy, ${a_p}_0=4$, confirms this trend, exhibiting a stronger impact on the cascade rate, both in the inertial range and at larger scales, close to the forcing scales. Furthermore, a Fourier-based numerical method, to compute exact laws in numerical simulations in the full $(\ell_\perp,\ell_\parallel)$ increment plane, is presented. 
\end{abstract}

\maketitle

\section{Introduction}

The turbulent state of fluids and plasmas (e.g., stars, nebula, stellar wind, interplanetary medium) is associated with the development of cascades of various quantities (such as energy or helicity), characterized by a constant transfer rate on a range of scales belonging to the inertial domain, where driving and dissipation are negligible. The total energy cascade rate can be used as a proxy of the energy dissipation, and thus, of the heating rate, in a turbulent medium. It can be estimated using third-order laws (also often called exact laws) that relate the cascade rate to the increments of the fluid moments, which are more easily measurable experimentally. Such laws can be obtained by assuming only the statistical homogeneity of the fluid. A more specific set of hypotheses (separation of injection and dissipation scales, statistical stationarity, high Reynolds numbers) is usually applied to isolate the nonlinear terms of the third-order law associated with the cascade process in the inertial range \citep{ferrand_-depth_2022}. Initiated by \citet{von_karman_statistical_1938} and \citet{kolmogorov_dissipation_1941} in the framework of incompressible hydrodynamics turbulence, the formalism has been extended to incompressible magnetohydrodynamics (MHD) \citep{politano_dynamical_1998,politano_von_1998}, Hall-MHD \citep{galtier_von_2008,banerjee_alternative_2017,hellinger_von_2018,ferrand_exact_2019}, and two-fluid flows \citep{andres_exact_2016,andres_von_2016}, and then to various compressible regimes characterized by different equations of state \citep{galtier_exact_2011,banerjee_exact_2013,banerjee_kolmogorov-like_2014,banerjee_scaling_2016,andres_alternative_2017,banerjee_exact_2017,andres_energy_2018,andres_exact_2018,banerjee_energy_2018,banerjee_scale--scale_2020,hellinger_cascade_2020,simon_general_2021}. Transfer rates of quantities different from the energy have also been considered in the literature \citep{soulard_cascade_2022,hellinger_spectral_2021,hellinger_cascade_2020,hellinger_scale_2021,hellinger_ion-scale_2022}.

 In the solar wind and other accessible natural plasmas, the third-order laws have been used to estimate the heating rate using in situ observations \citep{sorriso-valvo_observation_2007,andres_energy_2019,sorriso-valvo_turbulence-driven_2019,banerjee_scaling_2016, hadid_energy_2017,osman_proton_2013,carbone_scaling_2009,hadid_energy_2017,hadid_compressible_2018}. Since these laws grow in complexity when including more physical effects (e.g., compressibility, non-ideal terms in the Ohm's law, pressure anisotropy, etc.), one then must resort to numerical simulations to grasp the physics involved in each term, both for freely-decaying \citep{ferrand_compressible_2020,mininni_finite_2009,kritsuk_energy_2013, hellinger_von_2018,andres_energy_2018,ferrand_exact_2019,ferrand_-depth_2022} and driven turbulence \citep{ferrand_fluid_2021,jiang_energy_2023}.

 Recently, \citet{simon_general_2021,simon_exact_2022} proposed a generic derivation of the third-order law for the total energy, based on the internal-energy equation, and applied it to the Hall-MHD model with either isotropic or gyrotropic pressures. This work paves the way to more realistic studies of compressible turbulence in weakly collisional plasmas, such as those of the near-Earth space where the pressure is all but isotropic \citep{bale_magnetic_2009,hellinger_mirror_2017,hadid_compressible_2018}. In fluid description of plasmas, pressure anisotropy was first included by \citet{chew_boltzmann_1956}, within a MHD model, usually referred to as CGL, after the authors' names, or biadiabatic, because the heat fluxes are neglected. In the linear approximation, this model is known to permit the development of firehose and mirror-type instabilities \citep{hunana_introductory_2019,hunana_parallel_2017}. 
 In this paper, we propose to evaluate the cascade rate in driven-turbulence simulations of the CGL-Hall-MHD system for a proton-electron plasma, obtained by adding the Hall effect to the CGL model, with the aim to estimate quantitatively the effect of pressure anisotropy on the turbulent cascade, using the third-order law proposed in \citet{simon_exact_2022}. The study is performed on simulation data used in \citet{ferrand_fluid_2021}, complemented by a new simulation specifically designed for the present work. CGL-Hall-MHD is insufficient to provide a full description of the sub-ion range dynamics in space plasmas, where additional kinetic effects such as finite Larmor radius corrections, Landau damping, and cyclotron resonance may play an important role. However, it can be viewed as a relatively simple model to study the simultaneous influence of complex phenomena, such as compressibility, dispersion, and pressure anisotropy on the energy cascade in the MHD and the subionic ranges of scales.

 The paper is organized as follows. Section \ref{subsec:EL} recalls the main formulas derived in \citet{simon_exact_2022} that are used in the rest of the paper. Section \ref{sec:model} provides a brief description of the numerical code and of the simulation setups. Section \ref{sec:method} describes the Fourier-based method used to evaluate the third-order law in the numerical simulations. Section \ref{sec:results} summarizes the main results of the study, which are discussed in Section \ref{sec:disc}. 

\section{Third-order law for Hall-MHD turbulence with biadiabatic ions and isothermal electrons} \label{subsec:EL}

 The third-order laws for turbulent plasma described within the MHD and Hall-MHD approximations with a biadiabatic closure are derived in \cite{simon_exact_2022}. However, they should be extended to the regime of isothermal electrons considered in the simulations used in the present study.

 The CGL-Hall-MHD simulations at hand are based on the following evolution equations for the plasma density $\rho$, the fluid velocity $\boldsymbol{v}$, the magnetic field $\boldsymbol{B}$, the parallel $p_{\parallel}$ and perpendicular $p_{\perp}$ components of the gyrotropic ion pressure relatively to the local magnetic field, and the total (electron and ion) pressure tensor $ \overline{\boldsymbol{P}}$, assuming an isothermal electronic pressure $p_e$,
 
 \begin{eqnarray}\left\{
 \begin{split}
 \partial_t & \rho + \bnabla \bcdot (\rho \boldsymbol{v}) = - \nu_{\rho} \Delta_\alpha^4 \rho ,\\
 \partial_t & \boldsymbol{v} + \boldsymbol{v} \cdot \bnabla \boldsymbol{v} + \frac{1}{\rho} \bnabla \bcdot \overline{\boldsymbol{P}} - \frac{\boldsymbol{j}}{\rho} \times \boldsymbol{B} = \boldsymbol{f} - \nu_{v} \Delta_\alpha^4 \boldsymbol{v} ,\\
 \partial_t & {\boldsymbol{B}} - \bnabla \times (\boldsymbol{v} \times \boldsymbol{B}) + \frac{1}{R_i} \bnabla \times \left(\frac{\boldsymbol{j}}{\rho} \times \boldsymbol{B} \right) - \frac{1}{R_i} \bnabla \times \left(\frac{1}{\rho} \bnabla p_e \right) = - \nu_{B} \Delta_\alpha^4 \boldsymbol{B} ,\\
 \partial_t & p_{\parallel} + \bnabla \bcdot (p_{\parallel} \boldsymbol{v}) + 2p_{\parallel} \boldsymbol{b}\boldsymbol{b}:\bnabla\boldsymbol{v} = - \nu_{p} \Delta_\alpha^4 p_{\parallel} ,\\
 \partial_t & p_{\perp} + \bnabla \bcdot (p_{\perp} \boldsymbol{v}) + p_{\perp} \bnabla \bcdot \boldsymbol{v} - p_{\perp} \boldsymbol{b}\boldsymbol{b}:\nabla \boldsymbol{v} = - \nu_{p} \Delta_\alpha^4 p_{\perp}, \\
 \overline{\boldsymbol{P}} & = \frac{\beta_0}{2} \left(p_{\perp} + p_e \right)\overline{\boldsymbol{I}} + \frac{\beta_0}{2} (p_{\parallel} - p_{\perp}) \boldsymbol{b} \boldsymbol{b}, \\
 p_e &= \rho. 
 \end{split} 
 \right. \label{eq:model}
 \end{eqnarray}
Here, $\boldsymbol{j} =\bnabla \times \boldsymbol{B}$ is the current density, $\overline{\boldsymbol{I}}$ the identity tensor and ${\boldsymbol b} = \frac{\boldsymbol{B}}{|{\boldsymbol B}|}$ the unit vector in the direction of the local magnetic field. The density, magnetic field, velocity and pressures ($p_{\parallel}$, $p_{\perp}$ and $p_e$) are measured in units of the equilibrium density $\rho_0$, mean magnetic field $B_0$, Alfv\'en speed $c_{A} = \frac{B_0}{\sqrt{4 \pi \rho_0}}$, and equilibrium ion parallel pressure $p_{\parallel p_0}$, respectively. Furthermore, $R_i = \frac{L}{d_i} =\frac{L\sqrt{4 \pi e^2 \rho_0}}{ m_p c}$, where $L$ is the unit length, $d_i$ the ion inertial length, $e$ the elementary charge, $c$ the light speed and $m_p$ the proton mass, measures the importance of the Hall term relative to the nonlinearity. The ratio of the parallel ion pressure to the magnetic pressure is given by $\beta_0 /2 = \frac{p_{\parallel p_0}}{\rho_0 c^2_{A}}$. The time unit is $T = \frac{L}{c_{A}}$. The term $\boldsymbol{f}$ describes the injection of kinetic energy in the system, while the hyper-dissipative terms $\nu_{X} \Delta_\alpha^4 X$ induce small-scale dissipation of the generic field X. Note that the electronic pressure term, included in Ohm's law, cancels out for isentropic or isothermal electronic flows \citep{sahraoui_hamiltonian_2003}.

 The third-order law in the present setting can easily be obtained from the one derived in \cite{simon_exact_2022} for the two-point correlation function of total energy 
 \begin{equation}
 \mathcal{R} = \frac{1}{2}\lan\frac{1}{2}(\rho'+\rho)\boldsymbol{v}\cdot\boldsymbol{v'} + \frac{1}{2}(\rho'+\rho)\boldsymbol{v_A}\cdot\boldsymbol{v'_A} + \rho u' + \rho' u\ran, 
 \end{equation} 
 where
 $\boldsymbol{v_A} = \boldsymbol{B}/\sqrt{\rho}$ is the local Alfv\'en velocity,
 $\rho u = \frac{\beta_0}{4} \left(2 p_{\perp} + p_{\parallel} \right) + \rho_e u_e$
 is the density of internal energy, including the isothermal electronic contribution $\rho_e u_e = \frac{\beta_0}{2}\rho \ln \rho$. The operator $\lan\ran$ indicates an ensemble average, which is generally replaced by a spatial average, assuming ergodicity to hold. As usual, $\delta \varphi$ refers to the increment of a generic field $\varphi$ between the two positions $\boldsymbol{x}$ and $\boldsymbol{x'}=\boldsymbol{x} + \boldsymbol{\ell}$. Primed quantities are evaluated at the position $\boldsymbol{x'}$. Based on the sole assumption of statistical homogeneity, the "von Karman-Howarth-Monin (KHM) equation" that gives the evolution of $\mathcal{R}$ will read (we will go through the details gradually afterward): 
 \begin{equation}
 \partial_t \mathcal{R} + \varepsilon_{cgl} - \varepsilon_{F} - \varepsilon_{D} = 0. \label{eq:nstat_EL}
\end{equation} 
The non-linear term $\varepsilon_{cgl}$ is the total cascade rate and is expected to be equal to the mean injection rate $\varepsilon_{F}$ in the inertial range of scales of stationary turbulence, where the dissipation $\varepsilon_{D}$ rate is negligible and the temporal derivative $\partial_t \mathcal{R}$ vanishes. Its expression is given by

 \begin{equation}
 -4 \varepsilon_{cgl}(\boldsymbol{\ell}) = \bnabla_{\boldsymbol{\ell}} \cdot \boldsymbol{\mathcal{F}}(\boldsymbol{\ell}) + (\mathcal{S}(\boldsymbol{\ell}) + \mathcal{S}(-\boldsymbol{\ell})) + \varepsilon_{hall}(\boldsymbol{\ell})
 \end{equation}
with 
\begin{eqnarray}
 &&\left\{
 \begin{split}
 \boldsymbol{\mathcal{F}}(\boldsymbol{\ell}) &= \langle \left(\delta \left(\rho \boldsymbol{v}\right) \cdot \delta \boldsymbol{v} + \delta \left(\rho \boldsymbol{v_A}\right) \cdot \delta \boldsymbol{v_A} + 2 \delta \rho \delta u\right) \delta \boldsymbol{v} \rangle + \langle - \delta \rho \delta \left(\frac{\overline{\boldsymbol{P_*}}}{\rho}\right)\cdot \delta \boldsymbol{v} \rangle \, \\
 &-\langle\left( \delta \left(\rho \boldsymbol{v_A}\right) \cdot \delta \boldsymbol{v} + \delta \left(\rho \boldsymbol{v}\right) \cdot \delta \boldsymbol{v_A} \right) \delta \boldsymbol{v_A} \rangle \, ,\\
\mathcal{S} (\boldsymbol{\ell}) & = \langle \left(\rho \boldsymbol{v} \cdot \delta \boldsymbol{v} +\frac{1}{2} \rho \boldsymbol{v_A} \cdot \delta \boldsymbol{v_A} - \frac{1}{2}\boldsymbol{v_A} \cdot \delta \left(\rho \boldsymbol{v_A}\right) + 2 \rho \delta u\right)\bnabla' \bcdot \boldsymbol{v'} \rangle \\
&+ \langle \left( - 2 \rho \boldsymbol{v} \cdot \delta \boldsymbol{v_A} - \rho \boldsymbol{v_A} \cdot \delta \boldsymbol{v} + \delta (\rho \boldsymbol{v}) \cdot \boldsymbol{v_A} \right) \bnabla' \bcdot \boldsymbol{v'_A} \rangle \\
&+ \langle - 2\rho \delta \left(\frac{\overline{\boldsymbol{P}}}{\rho}\right) : \bnabla' \boldsymbol{v'} \rangle+ \langle\left(\delta \rho \frac{\overline{\boldsymbol{P_*}}}{\rho} \cdot \boldsymbol{v} - \rho \delta \left(\frac{\overline{\boldsymbol{P_*}}}{\rho}\right) \cdot \boldsymbol{v} \right)\cdot \frac{\bnabla' \rho' }{\rho'} \rangle, \\
\varepsilon_{hall}(\boldsymbol{\ell}) &= 2\frac{1}{R_i}\langle \overline{\boldsymbol{j} \times \boldsymbol{v_A}} \times \delta \boldsymbol{v_A} - \delta(\frac{\boldsymbol{j}}{\rho} \times \boldsymbol{v_A})\times \overline{\rho \boldsymbol{v_A}} \rangle \\
&+ \frac{1}{R_i} \langle \frac{1}{2} \left(\delta \rho \boldsymbol{v'_A} \cdot\boldsymbol{v_A} \right) \bnabla' \bcdot \frac{\boldsymbol{j'}}{\rho'} - \left( \delta \rho \frac{\boldsymbol{j'}}{\rho'} \cdot \boldsymbol{v_A}\right)\bnabla' \bcdot \boldsymbol{v'_A} \rangle ,
\end{split}
\right. \label{eq:extL}
\end{eqnarray}
 where $\overline{\boldsymbol{P_*}}$ is the total (thermal and magnetic) pressure tensor. The spatial derivative $\bnabla'$ is performed relative to the variable $\boldsymbol{x'}$, while 
 $\bnabla_{\boldsymbol{\ell}}$ is the spatial derivative operator in the scale space. 
 
 The contributions to the cascade rate (Eq. \eqref{eq:extL}) can be separated into the divergence of the flux density $\boldsymbol{\mathcal{F}}$ constituted of structure functions, source $\mathcal{S}$ and Hall $\varepsilon_{hall}$ terms. To analyze them, we define the isotropic, $p= \frac{\beta_0}{6} \left(2 p_{\perp} + p_{\parallel} \right) + \frac{\beta_0}{2} \rho $, and anisotropic, $\overline{\boldsymbol{\Pi}} = \overline{\boldsymbol{P}} - p \overline{\boldsymbol{I}}$, parts of the pressure. Therefore, the cascade rate \eqref{eq:extL} can be split into two parts, $\varepsilon_{iso}$ and $\varepsilon_{\overline{\boldsymbol{\Pi}}}$, the latter depending on the anisotropic part of the pressure tensor $\overline{\boldsymbol{\Pi}}$, while the former does not. One writes
\begin{equation}
-4 \varepsilon_{\overline{\boldsymbol{\Pi}}} = \bnabla_{\boldsymbol{\ell}} \cdot \boldsymbol{\mathcal{F}}_{\Pi} + \mathcal{S}_{\Pi1} + \mathcal{S}_{\Pi2} + \mathcal{S}_{\Pi3}
\end{equation}
with 
\begin{eqnarray}
\left\{
 \begin{split}
 \boldsymbol{\mathcal{F}}_{\Pi}(\boldsymbol{\ell}) &= - \langle \delta \rho \delta \left(\frac{\overline{\boldsymbol{\Pi}}}{\rho}\right)\cdot \delta \boldsymbol{v}\rangle ,\\
 \mathcal{S}_{\Pi1}(\boldsymbol{\ell})&= (\mathcal{S}(\boldsymbol{\ell}) + \mathcal{S}(-\boldsymbol{\ell}))_{\Pi1} = \langle 2\rho' \delta \left(\frac{\overline{\boldsymbol{\Pi}}}{\rho}\right) : \bnabla \boldsymbol{v} - 2\rho \delta \left(\frac{\overline{\boldsymbol{\Pi}}}{\rho}\right) : \bnabla' \boldsymbol{v'} \rangle,\\
 \mathcal{S}_{\Pi2}(\boldsymbol{\ell}) & = (\mathcal{S}(\boldsymbol{\ell}) + \mathcal{S}(-\boldsymbol{\ell}))_{\Pi3} = \langle \rho' \delta \left(\frac{\overline{\boldsymbol{\Pi}}}{\rho} \right)\cdot \boldsymbol{v'} \cdot \frac{\nabla \rho }{\rho} - \rho \delta \left(\frac{\overline{\boldsymbol{\Pi}}}{\rho} \right)\cdot \boldsymbol{v} \cdot \frac{\nabla' \rho' }{\rho'} \rangle ,\\
 \mathcal{S}_{\Pi3}(\boldsymbol{\ell}) &= (\mathcal{S}(\boldsymbol{\ell}) + \mathcal{S}(-\boldsymbol{\ell}))_{\Pi2} = \langle\delta \rho \frac{\overline{\boldsymbol{\Pi}}}{\rho} \cdot \boldsymbol{v} \cdot \frac{\bnabla' \rho' }{\rho'} - \delta \rho \frac{\overline{\boldsymbol{\Pi'}}}{\rho'} \cdot \boldsymbol{v'} \cdot \frac{\bnabla \rho }{\rho} \rangle .
\end{split}
\right. \label{eq:dtail_an}
\end{eqnarray}
 
 When computing third-order laws in numerical simulation, it is important to assess their level of accuracy. For this purpose, we will verify Eq. (\ref{eq:nstat_EL}), by computing $ \partial_t \mathcal{R}$, $\varepsilon_{F}$ and $\varepsilon_{D}$. In our simulations with the forcing and the hyper-dissipative terms, $\varepsilon_{F}$ and $\varepsilon_{D}$ are given by
\begin{eqnarray}
 \varepsilon_{F} &=& \frac{1}{4} \lan \left(\rho + \rho'\right) \left(\boldsymbol{v} \cdot \boldsymbol{f'} + \boldsymbol{v'} \cdot \boldsymbol{f}\right)\ran ,\\
 \varepsilon_{D} &=& - \frac{\nu_{v}}{4} \lan \left(\rho + \rho'\right) \left(\boldsymbol{v} \cdot \Delta'_\alpha{}^4 \boldsymbol{v'} + \boldsymbol{v'} \cdot \Delta_\alpha^4 \boldsymbol{v} \right)\ran - \frac{\nu_{\rho}}{4} \lan \left(\Delta_\alpha^4 \rho + \Delta'_\alpha{}^4 \rho'\right) \boldsymbol{v} \cdot \boldsymbol{v'} \ran \nonumber \\
 && - \frac{\nu_{B}}{4} \lan \left(\rho + \rho'\right) \left(\frac{1}{\sqrt{\rho'}}\boldsymbol{v_A} \cdot \Delta'_\alpha{}^4 \left(\sqrt{\rho'}\boldsymbol{v'_A}\right) + \frac{1}{\sqrt{\rho}} \boldsymbol{v'_A} \cdot \Delta_\alpha^4 \left(\sqrt{\rho}\boldsymbol{v_A}\right) \right)\ran \nonumber \\
 && + \frac{\nu_{\rho}}{8} \lan \left(\rho - \rho'\right) \left(\frac{\Delta_\alpha^4 \rho}{\rho} - \frac{\Delta'_\alpha{}^4 \rho'}{\rho'}\right) \boldsymbol{v_A} \cdot \boldsymbol{v'_A} \ran \nonumber \\
 && - \frac{\nu_{p} \beta_0}{8} \lan \frac{\rho}{\rho'} \Delta'_\alpha{}^4 \left(2 p'_{\perp} + p'_{\parallel}\right) + \frac{\rho'}{\rho} \Delta_\alpha^4 \left(2 p_{\perp} + p_{\parallel}\right)\ran \nonumber \\
 && - \frac{\nu_{\rho} \beta_0}{8} \lan \left(\frac{\Delta_\alpha^4 \rho}{\rho} - \frac{\Delta'_\alpha{}^4 \rho'}{\rho'}\right) \left(\frac{\rho}{\rho'} \left(2 p'_{\perp} + p'_{\parallel}\right) - \frac{\rho'}{\rho}\left(2 p_{\perp} + p_{\parallel}\right)\right) \ran \nonumber \\
 && - \frac{\nu_{\rho} \beta_0}{4} \lan \left( \frac{\rho'}{\rho} + \ln \rho'\right) \Delta_\alpha^4 \rho + \left(\frac{\rho}{\rho'} + \ln \rho \right)\Delta'_\alpha{}^4 \rho' \ran .
\end{eqnarray}

In fact, when computed numerically, the left-hand side of Eq. \eqref{eq:nstat_EL} is not strictly zero, but amounts to a value $\zeta$ that defines the numerical errors inherent to the computation of the various terms.

\section{Numerical setup} \label{sec:model}

\subsection{Simulation model} 
The numerical code used here is based on Eq. \eqref{eq:model}. Assuming periodic boundary conditions, the Cartesian three-dimensional space variables $\{x,y,z\}$ are discretized via a Fourier pseudo-spectral method, where most of the aliasing is suppressed by spectral truncation at $2/3$ of the maximal wavenumber in each direction. The time stepping is performed with a third-order low-storage Runge-Kutta scheme, with a prescribed time step $\delta t$. 

In the driven turbulence simulations considered here, a Langevin forcing $\boldsymbol{f}$ is added in the momentum equation (see Eq. \eqref{eq:model}). In Fourier space, this forcing appears as a Dirac distribution at the smallest wavevectors of the simulation domain and reflects the injection of balanced (in the sense that the same amount of energy is carried by forward and backward propagating
waves) dispersive Alfvén waves with amplitudes $A_f$, random phases, and a wavevector making an angle $\theta_i$ with the mean magnetic field. For two simulations (CGL1 and CGL2), forced in the MHD scales, the injected waves are close to MHD Alfvén waves.
After some time, a quasi-stationary regime is established in which the sum of the perpendicular kinetic and magnetic energies fluctuates around a prescribed level.
The forcing is turned on when the sum of the energies of the perpendicular kinetic and magnetic fluctuations reaches a floor level given by $93 \%$ of the prescribed maximal energy $E_{max}$, and is turned off when this latter level is reached. The internal energy $u$ can, however, continue to increase, with $p_\perp$ increasing and $p_\|$ decreasing (not shown), but on a typical time scale
longer than the characteristic time of the turbulence dynamics. In some runs (e.g. CGL3), the averaged pressures reach
a quasi-stationary level, while in others (e.g. CGL3b), they continue to evolve, but, due to
the separation in the time scales, the analysis of the transfer rate can be carried out safely.

Hyper-dissipation in the form of terms depending on $\Delta^4_\alpha$ and a coefficient is also added in the dynamical equations (Eq. \eqref{eq:model}) to smooth the solution at the smallest scales. The coefficients are adjusted to optimize the scale separation between injection and dissipation. Since the simulations develop spatial anisotropies with respect to the mean magnetic field taken in the $z$ direction, an anisotropy parameter $\alpha$ is introduced in the Laplacian operator $\Delta_\alpha = \partial^2_x+\partial^2_y+\alpha \partial^2_z$. 

\begin{table}
\begin{center}
\begin{tabular}{ r|c|c|c|c|c|c } 
Name & CGL1 & CGL2 & CGL3 & CGL3B & CGL5 & CGL6 \\
\hline
Resolution & $512^3$ & $512^3$ & $512^2 \times 1024$ & $512^2 \times 1024$ & $512^2 \times 1024$ & $512^2 \times 1024$ \\
\hline
$k^{(0)}_{\perp}d_i$ & 0.045 & 0.045 & 0.5 & 0.5 & 0.147 & 0.147 \\
$\theta_i$ & 83° & 75° & 75° & 75° & 75° & 75° \\
$E_{max}$ & $1.6 \times 10^{-2}$ & $1.6 \times 10^{-2}$ & $4.5 \times 10^{-2}$ & $1.125 \times 10^{-2}$ & $1.6 \times 10^{-2}$ & $1.6 \times 10^{-2}$ \\
$A_f$ & $1.0 \times 10^{-3}$ & $1.0 \times 10^{-3}$ & $8.0 \times 10^{-3}$ & $4.0 \times 10^{-3}$ & $3.0 \times 10^{-3}$ & $3.0 \times 10^{-3}$ \\
\hline
$\nu_{v} = \nu_{B}$ & $7.35 \times 10^{-8}$ & $7.35 \times 10^{-8}$ & $4.0 \times 10^{-14}$ & $1.0 \times 10^{-14}$ & $3.0 \times 10^{-11}$ & $3.0 \times 10^{-11}$ \\
$\nu_{\rho}$ & 0 & 0 & $1.6 \times 10^{-14}$ & $1.0 \times 10^{-14}$ & 0 & 0 \\ 
$\nu_{p}$ & $7.35 \times 10^{-9}$ & $7.35 \times 10^{-9}$ & $1.6 \times 10^{-14}$ & $1.0 \times 10^{-14}$ & $3.0 \times 10^{-12}$ & $3.0 \times 10^{-12}$ \\
$\alpha$ & 80 & 10 & 2.5 & 2.5 & 6 & 5 \\
\hline
$\delta t$ & $6.25 \times 10^{-2}$ & $5 \times 10^{-2}$ & $2 \times 10^{-4}$ & $3 \times 10^{-4}$ & $5 \times 10^{-3}$ & $5 \times 10^{-3}$ \\
$t_I$ & 6700 & 12900 & 361 & 410 & 12905 & 2730 \\
\hline
$a_{p0}$ & 1 & 1 & 1 & 1 & 1 & 4 
\end{tabular}
\caption{Parameters of the various runs:
$k^{(0)}_{\perp}d_i$, $\theta_i$, $E_{max}$, $A_f$} are forcing parameters (injection wavenumber, angle, threshold, and amplitude, respectively), while $\nu_{v} = \nu_{B}$, $\nu_{\rho}$, $\nu_{p}$ are hyper-dissipation coefficients arising in the dynamical equations for the velocity, the magnetic field, the density, and the pressures, respectively. The anisotropy parameter $\alpha$ enters the definition of the anisotropic Laplacian used in the simulations. The parameters $\delta t$ and $t_I$ refer to the time step and to the time at which the analysis was performed, while $a_{p0}$ measures the initial pressure anisotropy.
\label{tab:setups}
\end{center}
\end{table}

A summary of the parameter setup for the reported simulations is given in Table \ref{tab:setups}. 
Except for run CGL1 for which the driving angle is $\theta = 83^\circ$, this angle is taken equal to $75^\circ$ in all the other simulations. While in the runs CGL1 and CGL2, the driving acts at a transverse wavenumber $k^{(0)}_{\perp}$ located in the MHD range ($k^{(0)}_{\perp} d_i= 0.045$), in runs CGL3 and CGL3b, it acts in the close Hall range ($k^{(0)}_{\perp} d_i= 0.5$), the difference between these two simulations originating from the prescribed energy level $E_{max}$ that is four times larger in the former. Run CGL5 is initialized as CGL6, except for the pressures, with the driving located in the MHD range close to the Hall range ($k^{(0)}_{\perp} d_i=0.145$). 
We can estimate a Reynolds number characterizing the prevalence of the non-linear process over the dissipative one as $R_e \sim \frac{v L^7}{\nu_v}$ because of the use of a $(-\Delta)^4$ hyper-dissipation, $L \sim 1/(k^{(0)}_{\perp}d_i)$ denoting the scale of energy injection. From the values of $\nu_v$ and $k^{(0)}_{\perp}d_i$, and knowing that the velocity amplitude is about $v \sim 0.1$ for all simulations, we get huge values from $R_e \sim 10^{14}$ to $R_e \sim 10^{24}$ according to the simulations, despite the moderate scale separation between injection and dissipation and thanks to the use of hyper-dissipation.

\begin{figure}[!h]
\center
\includegraphics[width=0.9\textwidth,trim = 0.5cm 1cm 0cm 3cm, clip]{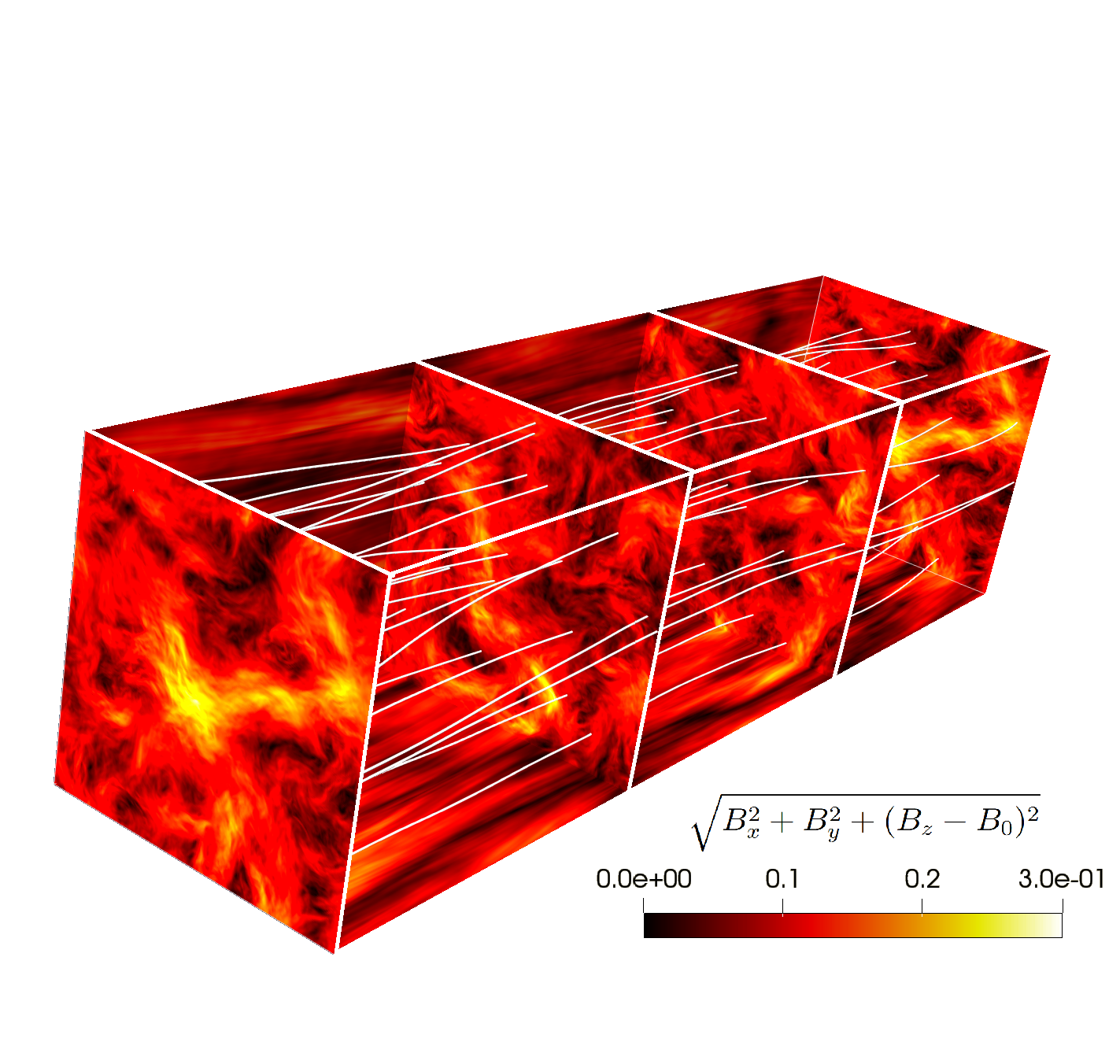}
\caption{3D rendering in the physical space of the magnetic field for CGL5. The color reflects the amplitude of the magnetic fluctuations $\sqrt{B_x^2+B_y^2+(B_z-B_0)^2}$, while the white lines are the global field lines.} 
\label{fig:3D}
\end{figure}

The initial conditions for the different fields are $\rho_0 = 1$, $\boldsymbol{v_0} = [0,0,0]$ and $\boldsymbol{B_0} = [0,0,1]$. The ion beta $\beta_{i0}$ and the dimensionless ion inertial length $d_i$ are initialized at 1. The pressure anisotropy $a_{p}=\frac{p_\perp}{p_\parallel}$ is initialized at $a_{p0}=1$ for all the runs, except CGL6 where it is set equal to 4. The runs CGL1, CGL2, and CGL3 were studied in \citet{ferrand_fluid_2021}, in a different framework. A three-dimensional rendering of the magnitude of the magnetic fluctuations together with a few total magnetic field lines is shown in Fig. \ref{fig:3D}.

\begin{figure}[!h]
\center
\includegraphics[width=0.5\textwidth,trim = 0cm 0cm 0cm 0cm, clip]{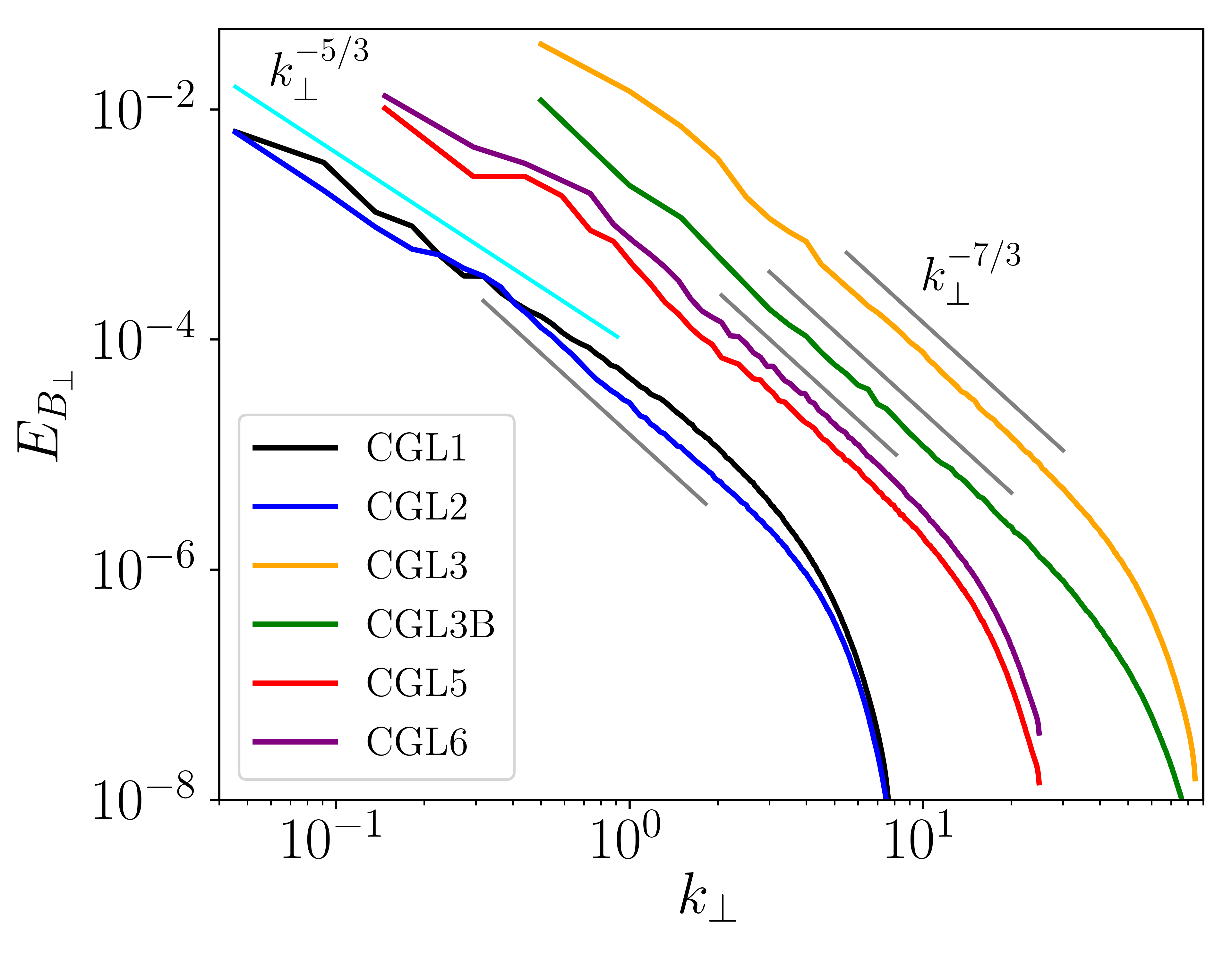}
\caption{Power spectra of the perpendicular magnetic field for the various simulations. The expected slope in the inertial (resp. Hall) range is indicated in cyan (resp. gray). The spectra of CGL1 and CGL2 are shifted down by a factor 20, for clarity.} 
\label{fig:Bperp_spectra}
\end{figure}

The data set used for each simulation is extracted at the time $t_I$ given in Table \ref{tab:setups}. At this time, the perpendicular magnetic energy spectrum has stabilized in a quasi-stationary regime and displays spectral exponent close to the expected values $-5/3$ and $-7/3$, signatures of an inertial range in the MHD and subion ranges of scales, respectively (See Fig. \ref{fig:Bperp_spectra}). These results indicate that the energy injected at the largest scales of the simulations by the forcing (that acts almost continuously) cascades through an inertial range extending on one to two decades, down to the smallest scales where the hyper-dissipation can act efficiently.

\subsection{Pseudo-spectral calculation of the structure functions} \label{sec:method} 

As previously mentioned, the expression of a cascade rate can be written in the general form 
$$-4 \varepsilon(\boldsymbol{\ell}) = \bnabla_{\boldsymbol{\ell}} \cdot \boldsymbol{\mathcal{F}}(\boldsymbol{\ell}) + ({\widetilde{\mathcal{S}}(}\boldsymbol{\ell}) + {\widetilde{\mathcal{S}}}(-\boldsymbol{\ell})),$$
where $\varepsilon$ is the cascade rate and ${\widetilde{\mathcal{S}}}$ collects all the terms not included in the divergence of ${\boldsymbol{\mathcal{F}}}$.

A straightforward method to compute it on the whole scale grid is to code a loop on all the possible separations $\boldsymbol{\ell}$ whose components in each direction are multiple of the mesh size in this direction, up to half the corresponding linear extension of the domain in this direction, to compute spatial derivatives with a four-point difference scheme for good accuracy, and to approximate the ensemble average by a spatial average over all the collocation points of the separation domain. 
However, such a procedure is extremely time- and memory-consuming. Consequently, in previous works \citep{ferrand_fluid_2021,ferrand_-depth_2022,andres_energy_2018}, the computation was performed on only a fraction of scales, associated with wisely selected directions in the separation space. Here, we introduce a different method, based on the equivalence between the two-point correlation function and a convolution product.
Thanks to this remark, we can compute the whole scale grid at once. Indeed, for two fields $A$ and $B$, the correlation product $\lan A'\cdot B\ran = \int A(\boldsymbol{x}+\boldsymbol{\ell}) B(\boldsymbol{x}) d\boldsymbol{x}$ can be viewed as the convolution product $A(\boldsymbol{\ell}) * B(-\boldsymbol{\ell}) = \int A({\boldsymbol x}) \cdot B(-(\boldsymbol{\ell}-{\boldsymbol x}) ) d{\boldsymbol x}$ that can easily be done in Fourier space as a simple product of the Fourier transform of $A$ and of the complex conjugate of the Fourier transform of $B$.
This protocol is possible for all terms in the source part ${\widetilde{\mathcal{S}}}$. For the terms entering $\boldsymbol{\mathcal{F}}$, the nonlinearity of which is higher than quadratic, the procedure is to be performed iteratively. The outcome is a 3D array in the scale space. 

Assuming that the scale space is axisymmetric with respect to the mean magnetic field and that all the considered statistical quantities are even in $\boldsymbol{\ell}$, 2D maps in function of $\ell_{\perp}$ and $\ell_{\parallel}>0$ are obtained by angular averaging in each perpendicular planes. These full 2D plots provide information about the cascade anisotropy in turbulence experiments \citep{lamriben_direct_2011} and numerical simulations \citep{manzini_local_2022}. To describe the cascade in the perpendicular direction, one may consider the 1D profiles, functions of $\ell_{\perp}$ only, which are obtained by restricting the considered quantity to the plane $\ell_{\parallel}=0$. This procedure mirrors the integration, in the Fourier space, over all the $k_\parallel$, to obtain the spectral distribution in $k_{\perp}$. Reciprocally, integration over all the parallel scales $\ell_\parallel$ is equivalent, in Fourier space, to consider the specific $k_\parallel=0$ plane. Note that, for the sake of simplicity, we keep using the same notation $\varepsilon$ for the 2D cascade rate $\varepsilon(\ell_\perp,\ell_\parallel)$ and the cascade rate in the perpendicular direction (in brief, perpendicular cascade rate) $\varepsilon(\ell_\perp)=\varepsilon(\ell_\perp,\ell_\parallel=0)$, as the context prevents any ambiguity. 

The time derivative needed to compute $\zeta$ is obtained using a two-point finite-difference scheme around the selected time $t_I$ (see Table \ref{tab:setups}).

\section{Results} \label{sec:results}

In this section, the results of the computation of the third-order law on the runs CGL1-CGL6 are analyzed. We recall that CGL1 and CGL2 cover mostly MHD scales, while CGL3 targets the smaller (Hall) scales. CGL3B is similar to CGL3 with a lower energy content, CGL5 was designed to cover the transition between the MHD and Hall ranges, by prescribing a forcing at an intermediate scale, CGL6 is similar to CGL5 except for the initial pressure anisotropy. Combined together, these simulations allow us to cover a range of scales extending on more than 2.5 decades, and spanning both the MHD and the subion scales. We start our analysis by considering the cascade in the perpendicular direction, before turning to the full 2D maps. 

\subsection{Global contributions to the KHM equation } \label{subsec:res_KHM}
\begin{figure}
 \begin{center}
 \includegraphics[width=0.49\textwidth,trim = 0.5cm 2.3cm 0.5cm 0.7cm, clip]{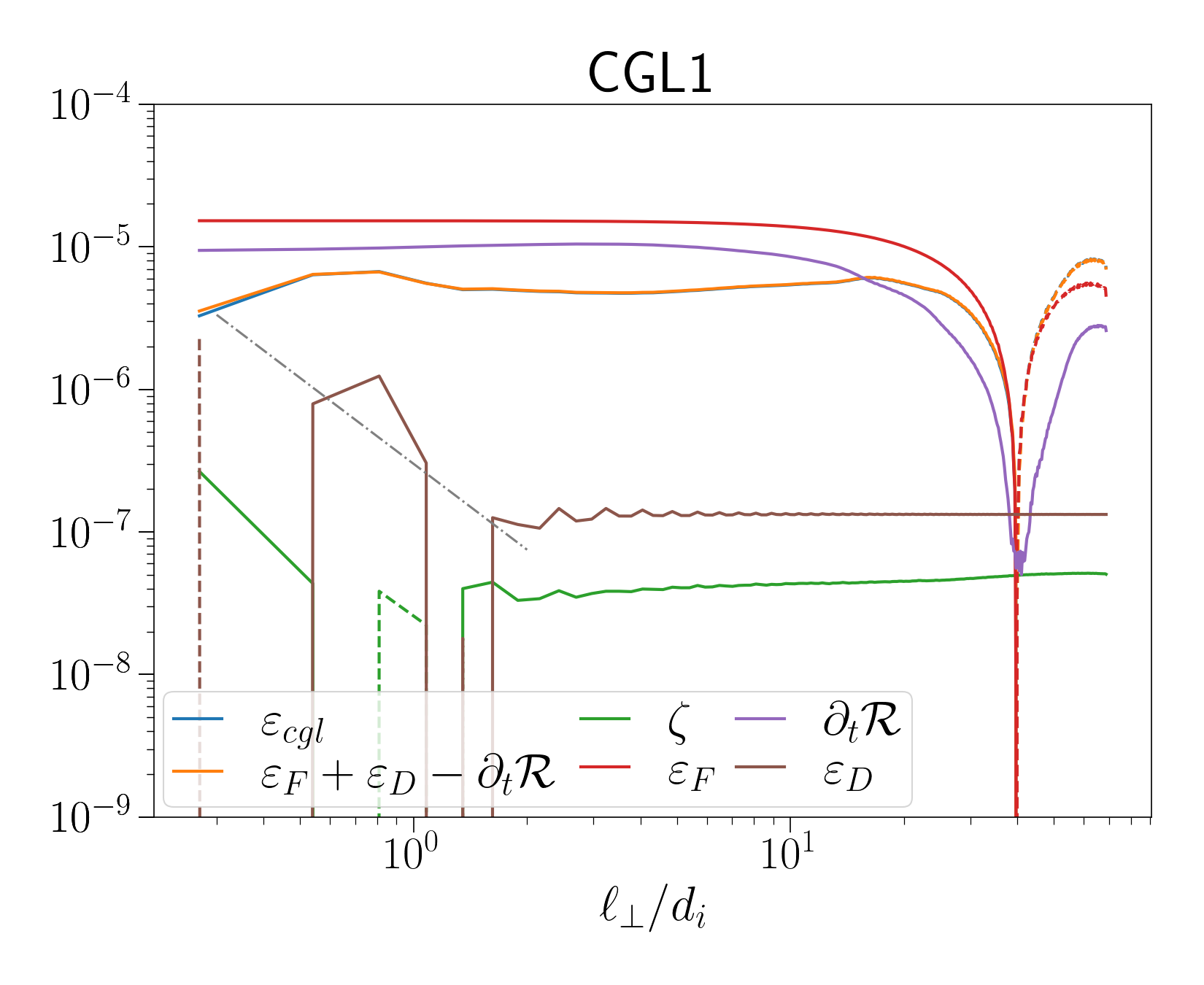} 
 \includegraphics[width=0.49\textwidth,trim = 0.5cm 2.3cm 0.5cm 0.7cm, clip]{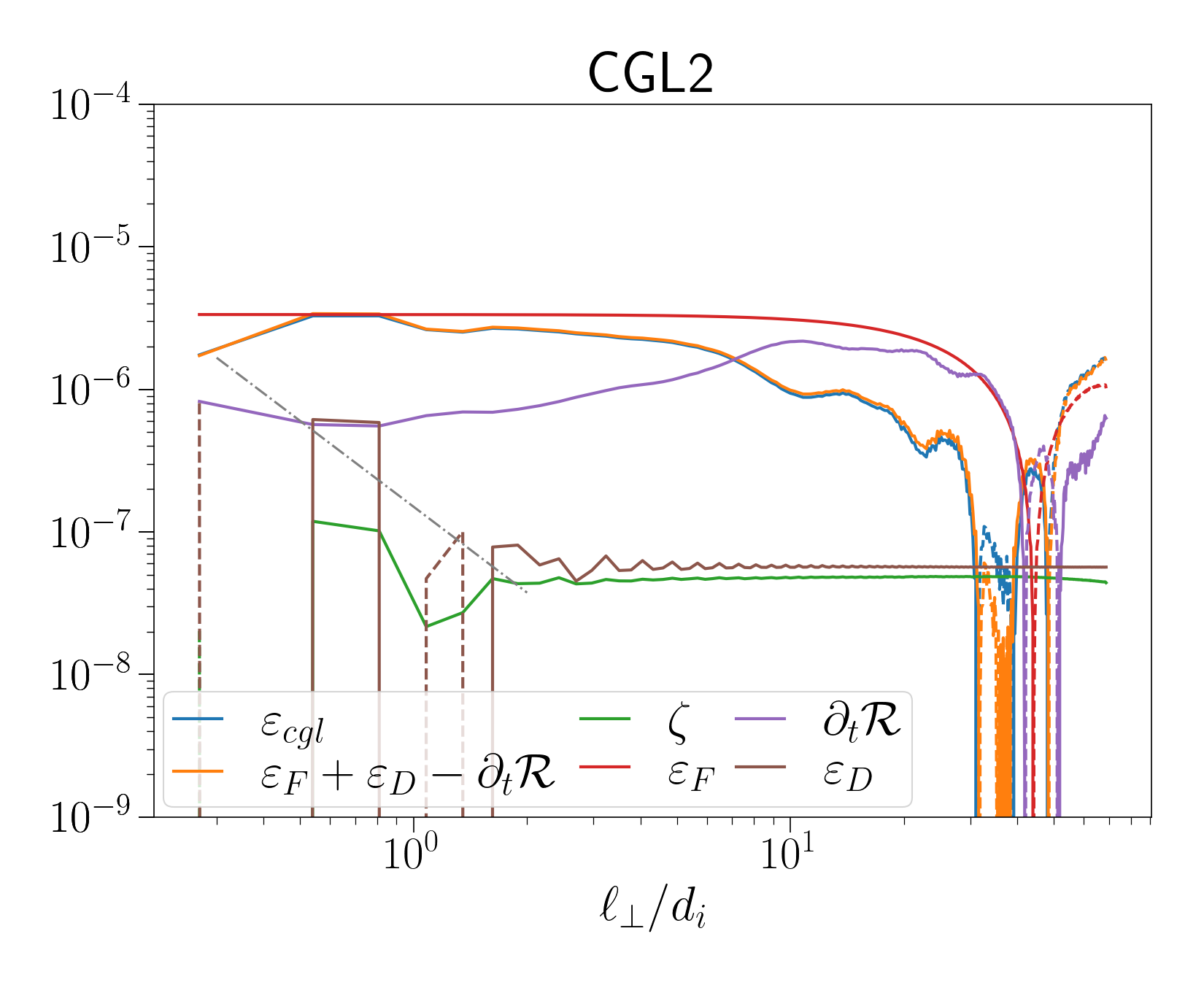} \\
 \includegraphics[width=0.49\textwidth,trim = 0.5cm 1.2cm 0.5cm 0.7cm, clip]{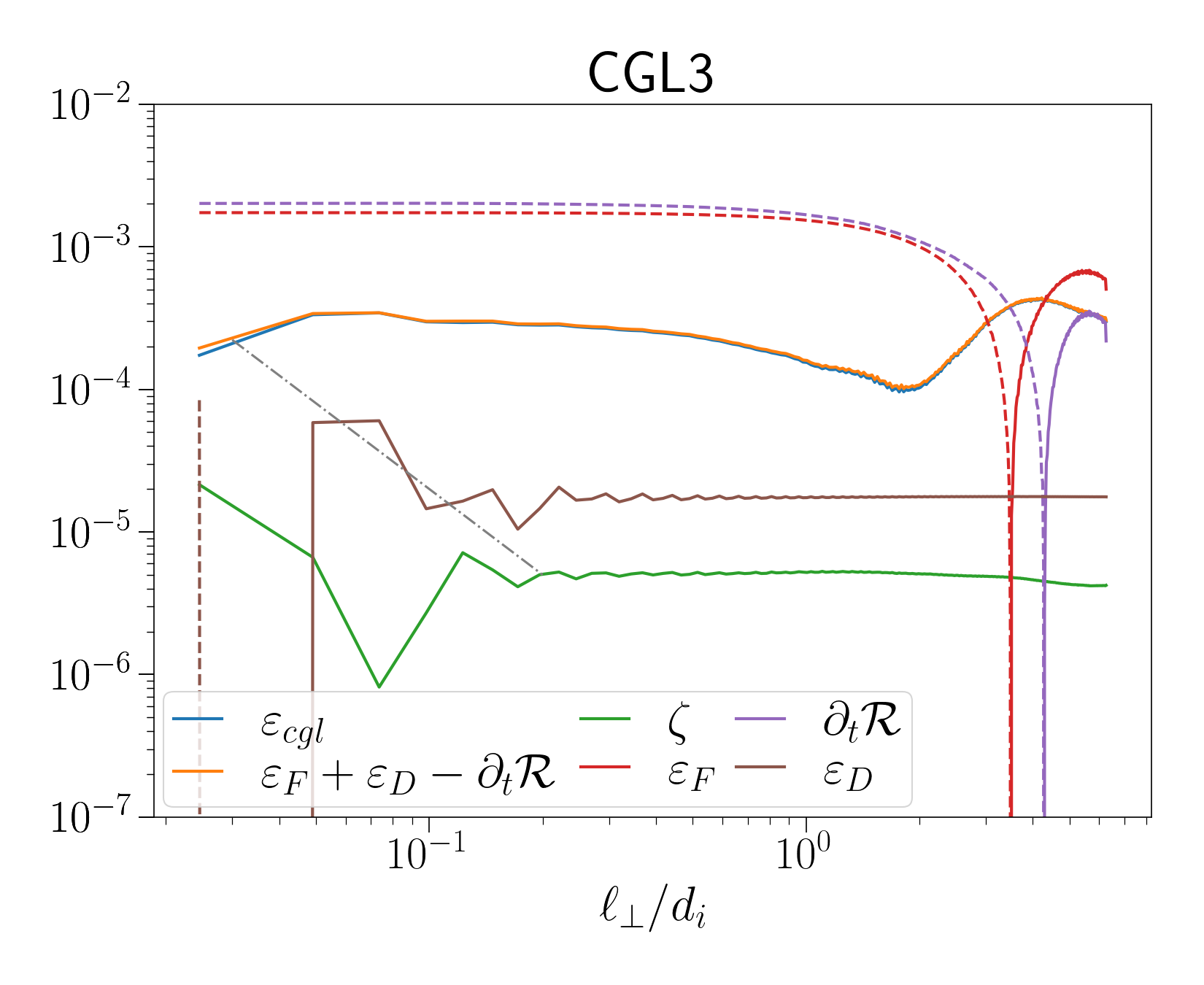} 
 \includegraphics[width=0.49\textwidth,trim = 0.5cm 1.2cm 0.5cm 0.7cm, clip]{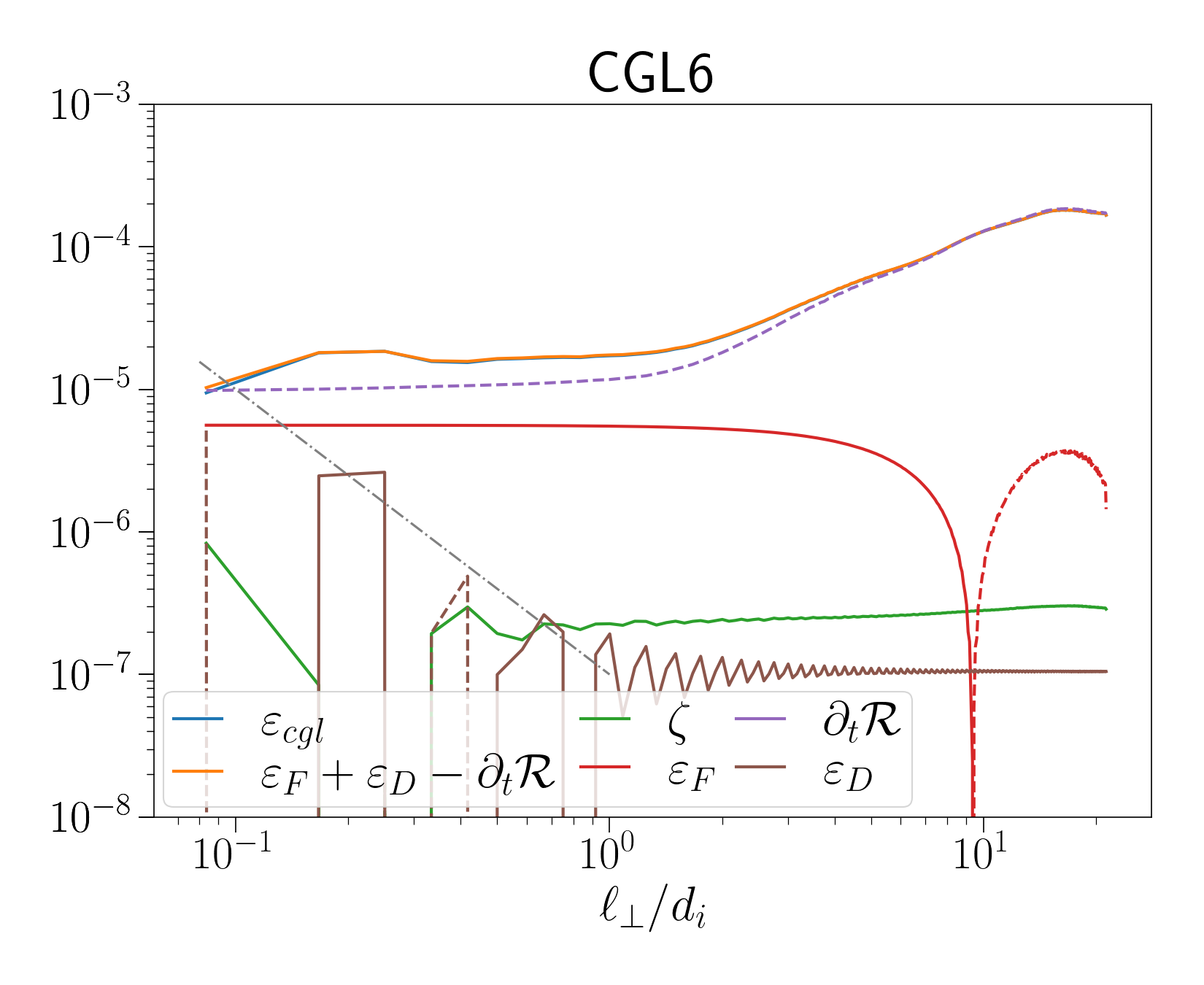} 
 \end{center}
 \caption{1D perpendicular profiles of the various terms of Eq. \eqref{eq:nstat_EL} for CGL1, CGL2, CGL3 and CGL6.
 Here and in the similar graphs presented in the following, solid lines refer to positive quantities, and dashed lines to the absolute value of negative quantities. Note that the blue line is almost indistinguishable from the orange one, consistent with Eq. \eqref{eq:nstat_EL} and the low level of $\zeta$. The gray dot-dashed line corresponds to a $-2$ slope.} 
 \label{fig:KHM}
 \end{figure}
 
The 1D perpendicular profiles ($\ell_\parallel= 0$) of the various terms entering Eq. (\ref{eq:nstat_EL}) and of the numerical error $\zeta$ are shown for CGL1, CGL2, CGL3 and CGL6 in Fig. \ref{fig:KHM}. In the increment space, the forcing term $\varepsilon_F$ is nearly constant down to the smallest scales, consistent with a forcing in the form of a Dirac distribution in Fourier space. 
The term $\partial_t \mathcal{R}$ reflects the instantaneous time fluctuations that, as in decaying turbulence \citep{ferrand_-depth_2022}, do not prevent the establishment of an extended inertial range.
The properties of $\varepsilon_F$ and $\partial_t \mathcal{R}$ differ in the various simulations: for CGL1 and CGL2 (and CGL3B --not shown), both terms are positive at most of the scales, for CGL3 (and CGL5 --not shown), they are mostly negative, while for CGL6, they have opposite signs. Furthermore, the instantaneous behavior of $\partial_t \mathcal{R}$ seems to compensate that of the forcing in order to keep the inertial range quasi-steady in time (not shown). Part of the injected energy is, in fact, directly converted to internal energy \citep{ferrand_fluid_2021}, which leads to enhancement and fluctuations of $\varepsilon_{cgl}$ at the largest scales, as demonstrated in the following.

The dissipation term $\varepsilon_D$ has an envelope consistent with a power law of exponent $-2$, at the smallest scales. This reflects a divergence of the two-point correlation function of a variable that has a power-law spectrum $k^{-3}$ or steeper, which is the case of the spectra shown in Fig. \ref{fig:Bperp_spectra} that steepen significantly at large wavenumber because of hyper-dissipation. This issue can be overcome by using higher-order correlation functions \citep{cho_simulations_2009}. We will not further discuss the dissipation term in this paper considering that it represents about $1\%$ of $\varepsilon_{cgl}$ in the inertial range, and thus does not significantly affect the cascade rate, but at the smallest scales.

We observe that $\varepsilon_{cgl}$ (blue) coincides with $- \partial_t \mathcal{R} + \varepsilon_{F} + \varepsilon_{D} $ (purple) with an error $\zeta$ (brown) between $10^{-7}$ and $10^{-8}$ for CGL1 and CGL2, between $10^{-5}$ and $10^{-6}$ for CGL3 and between $10^{-6}$ and $10^{-7}$ for CGL6. The uncertainty level is thus about $1\%$ to $10\%$ of $\varepsilon_{cgl}$, which makes it possible to analyze the various components of $\varepsilon_{cgl}$ in these simulations in a relatively reliable way.
 
\begin{figure}
 \center
 \includegraphics[width=0.506\textwidth,trim = 0cm 1.2cm 1.2cm 1.7cm, clip]{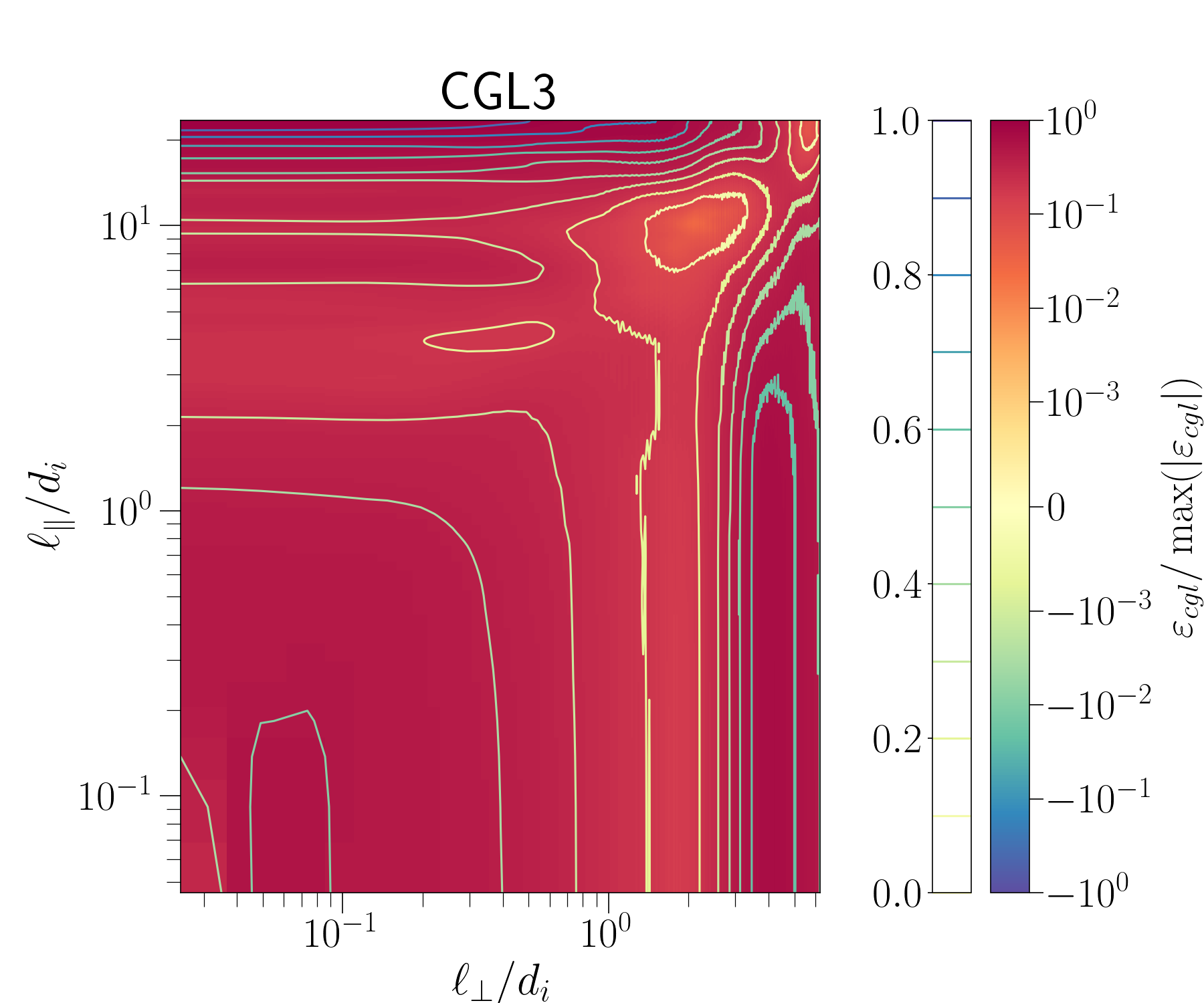} 
 \includegraphics[width=0.486\textwidth,trim = 2cm 1.2cm 0cm 1.7cm, clip]{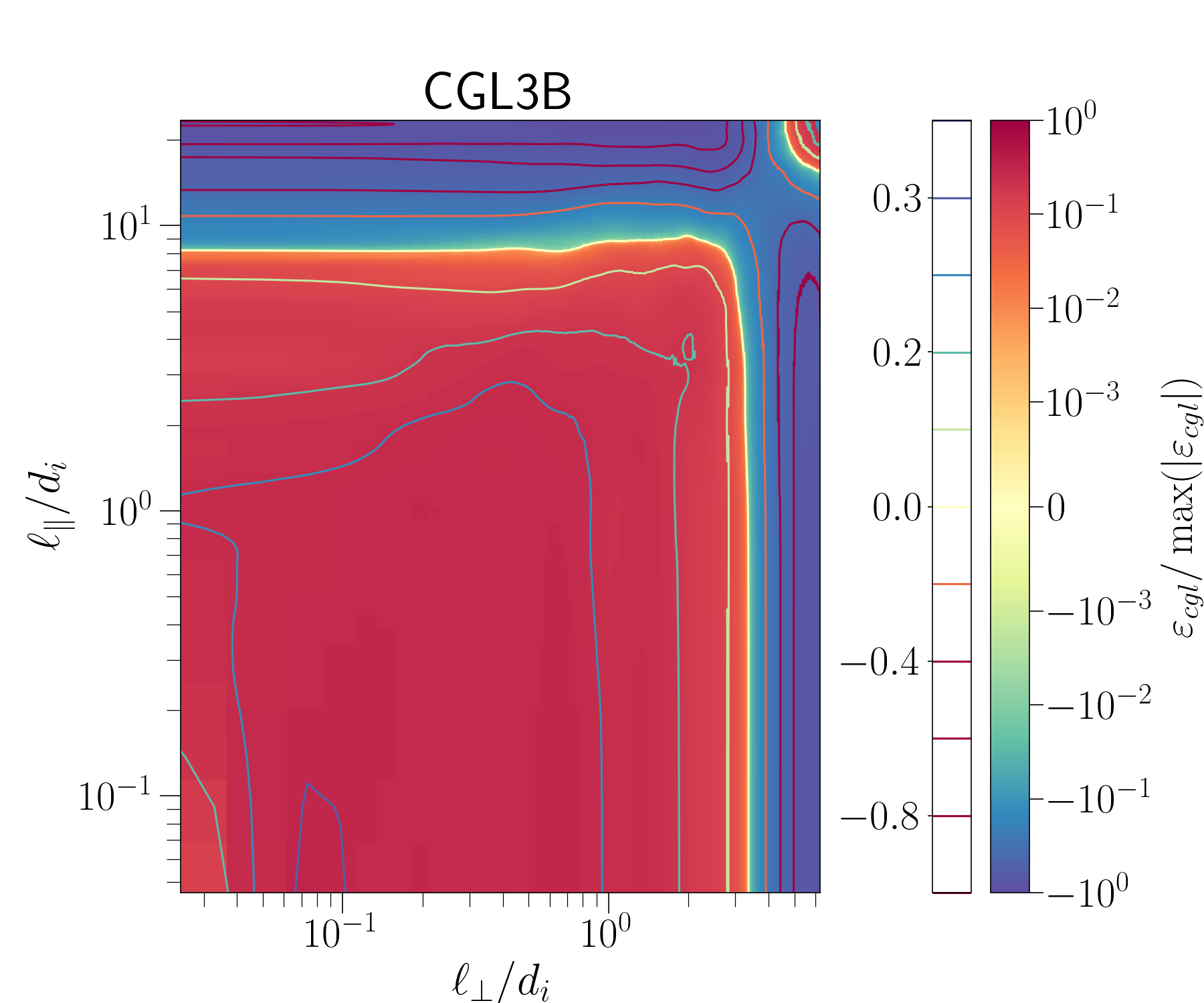} 
 \includegraphics[width=0.506\textwidth,trim = 0cm 0cm 1.2cm 1.7cm, clip]{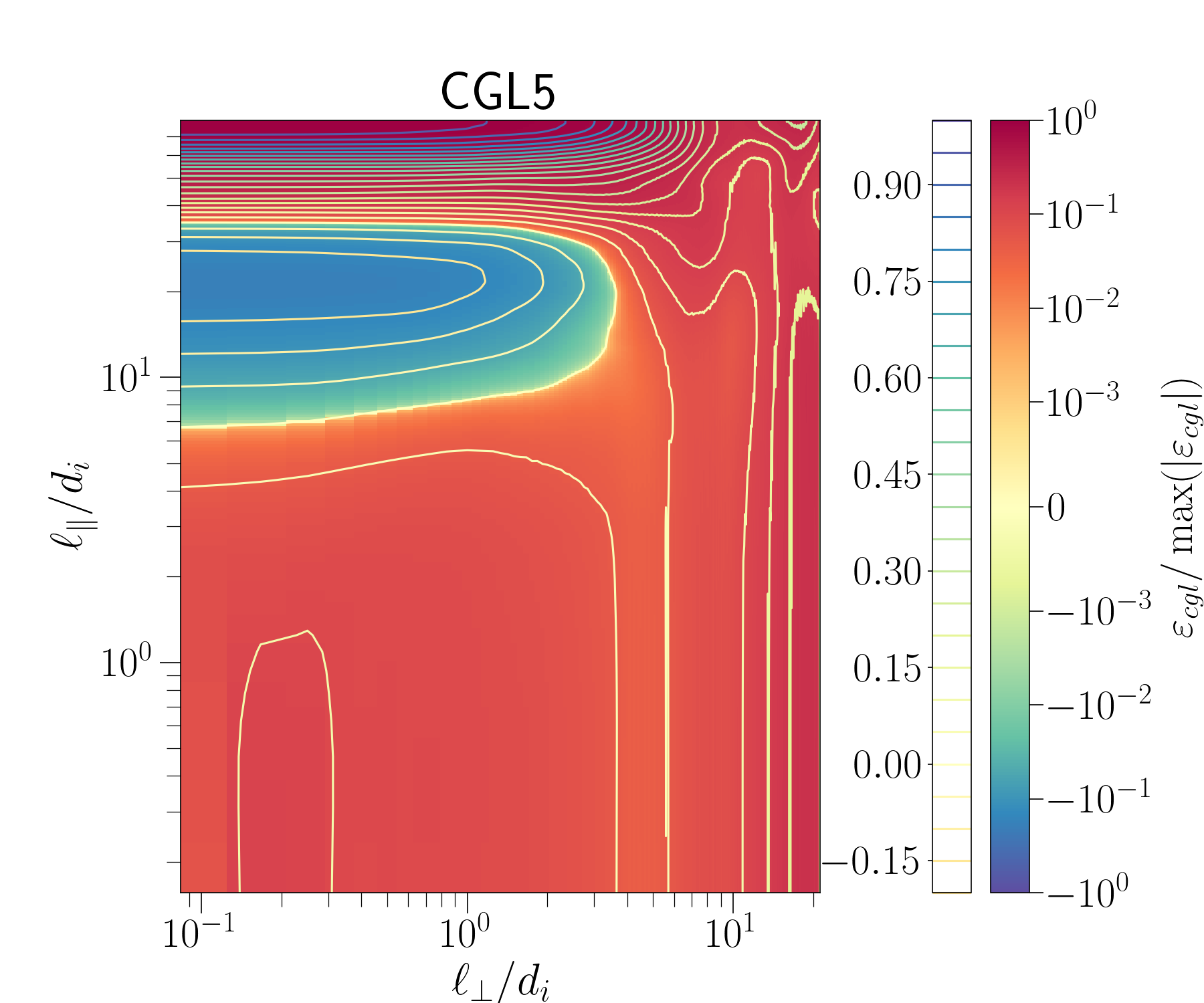} 
 \includegraphics[width=0.486\textwidth,trim = 2.5cm 0cm 0cm 1.7cm, clip]{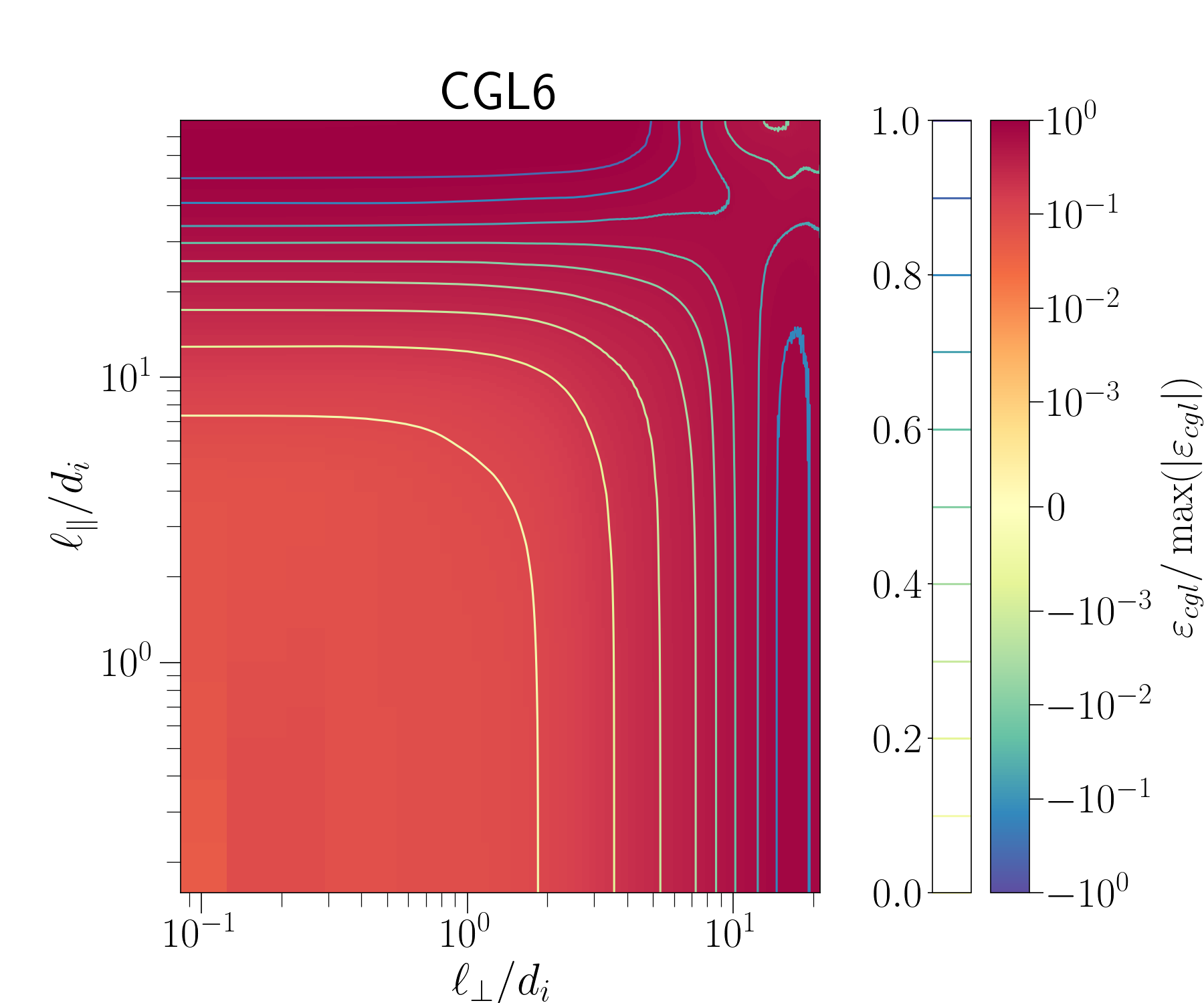} 
 \caption{2D maps of $\varepsilon_{cgl}/\max(|\varepsilon_{cgl}|)$ for CGL3 (top-left), CGL3B (top-right), CGL5 (bottom-left) and CGL6 (bottom-right). An orange-red background indicates a positive non-linear transfer and a green-blue background, a negative one. Contour lines are based on the linear scale displayed on the left of the color bar.} 
 \label{fig:epsCGL}
 \end{figure}

Figure \ref{fig:epsCGL} shows the resulting 2D maps of the normalized total cascade rate $\varepsilon_{cgl}$. The background colors indicate the variation of $\varepsilon_{cgl}$ with the positive values in warm colors and the negative ones in cold colors. Superimposed are a few contour lines. As visible on the perpendicular cascade rate (Fig. \ref{fig:KHM}), $\varepsilon_{cgl}$ decreases slightly at small scales and varies significantly at the large ones, while displaying a positive plateau at the intermediate $(\ell_{\perp},\ell_{\parallel})$ scales. This plateau defines the inertial range. We remark that in the $(\ell_{\perp},\ell_{\parallel})$ plane, the inertial range covers a domain whose shape is close to a square because of the logarithmic axes. One also notices that the large-scale variations of $\varepsilon_{cgl}$ are quite different in the various simulations: in CGL3B and CGL5 the sign changes, while in CGL3 and CGL6 $\varepsilon_{cgl}$ is positive with an enhancement in the forcing range and, for CGL3, a slight decrease at the transition between the inertial and the forcing scales. The behavior of the cascade rate in CGL5 is similar to that in CGL3B in the perpendicular direction, including the sign variations. A possible interpretation of these features is presented in the following, by analyzing the different components of $\varepsilon_{cgl}$.

\subsection{Isotropic and anisotropic pressure-induced contributions, $\varepsilon_{iso}$ and $\varepsilon_{\overline{\boldsymbol{\Pi}}}$, to the total cascade rate $\varepsilon_{cgl}$}

\begin{figure}
\center
\includegraphics[width=0.49\textwidth,trim = 1cm 2.5cm 1cm 1cm, clip]{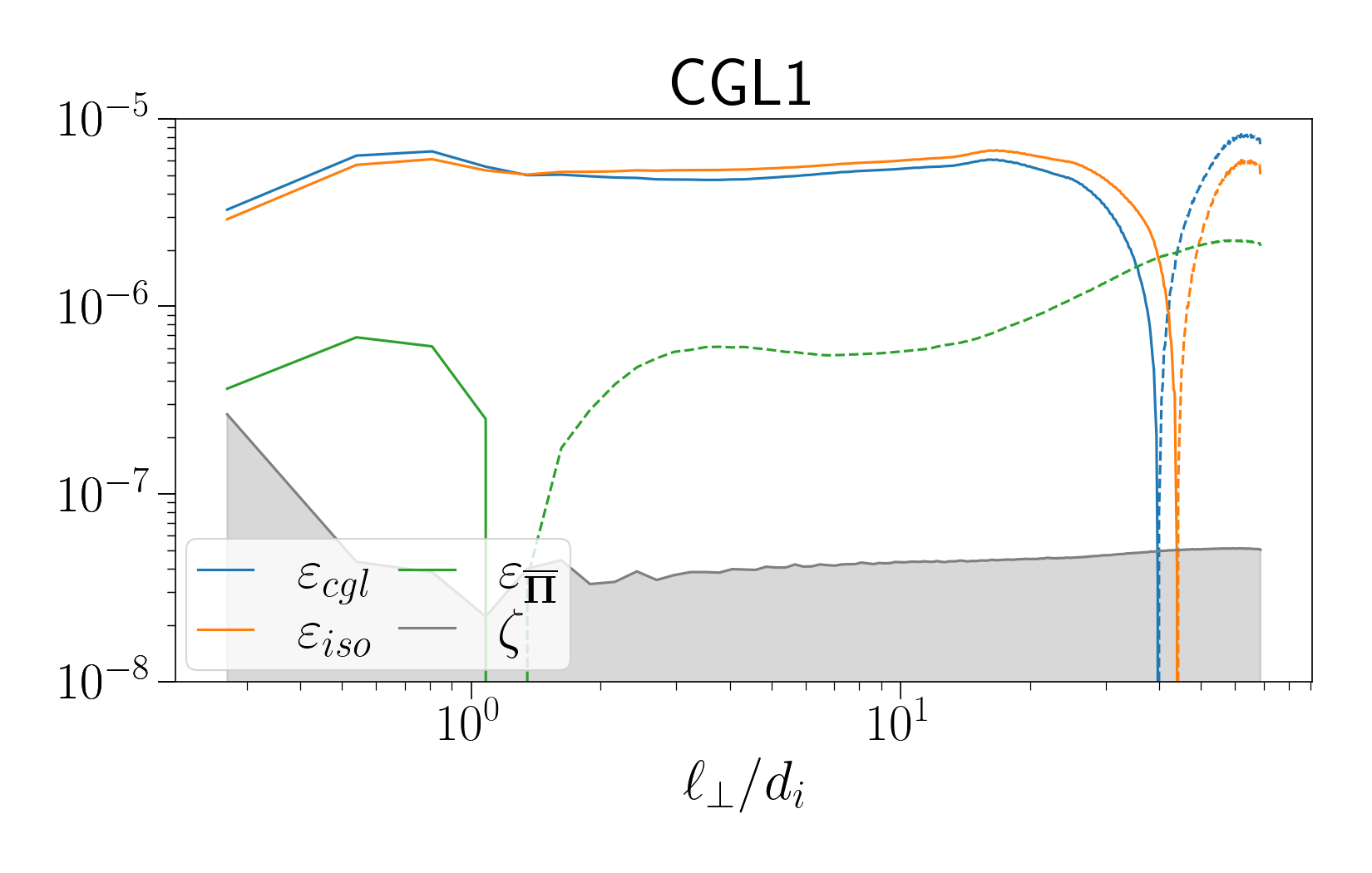}
\includegraphics[width=0.49\textwidth,trim = 1cm 2.5cm 1cm 1cm, clip]{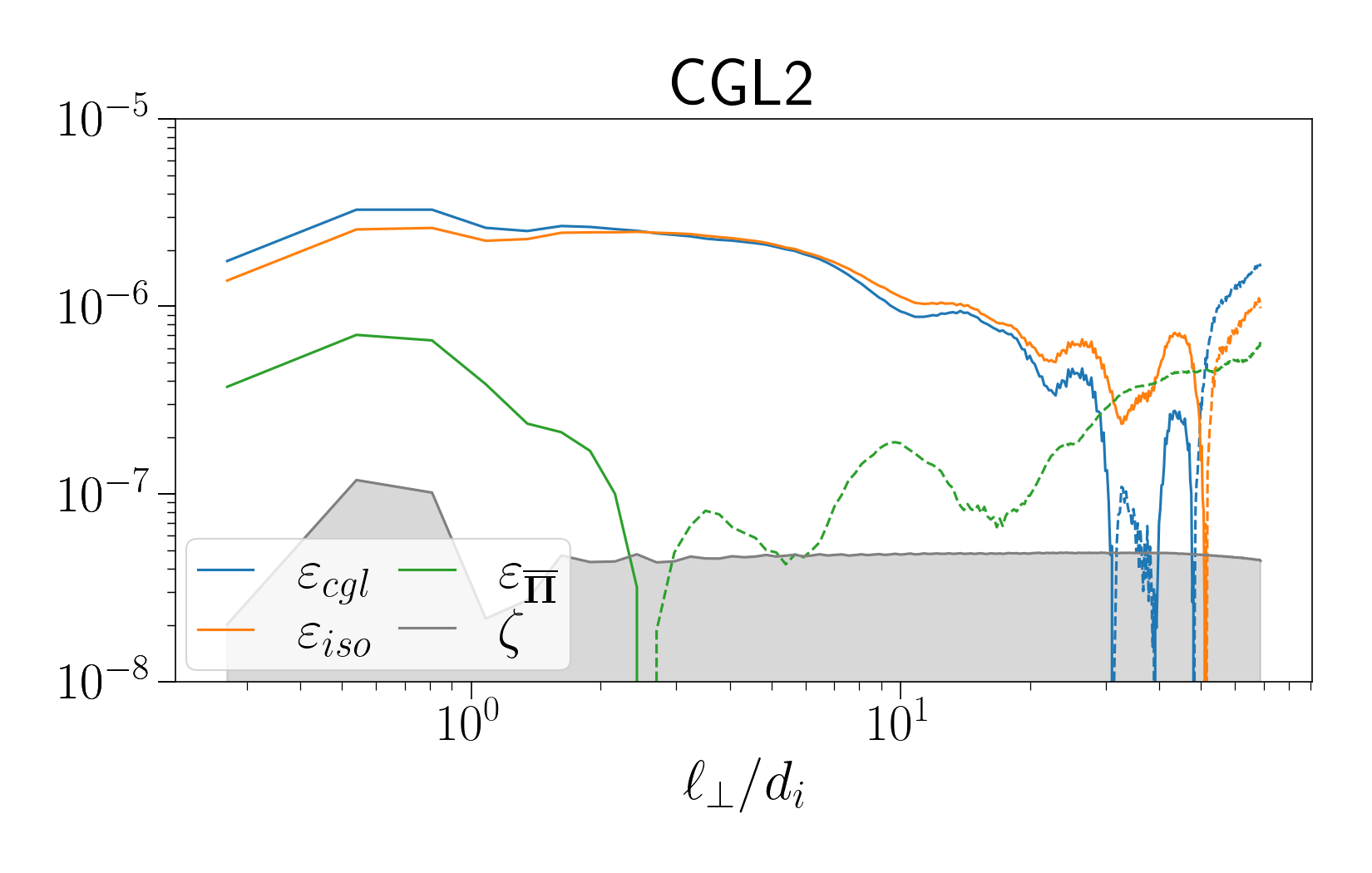}
\includegraphics[width=0.49\textwidth,trim = 1cm 2.5cm 1cm 1cm, clip]{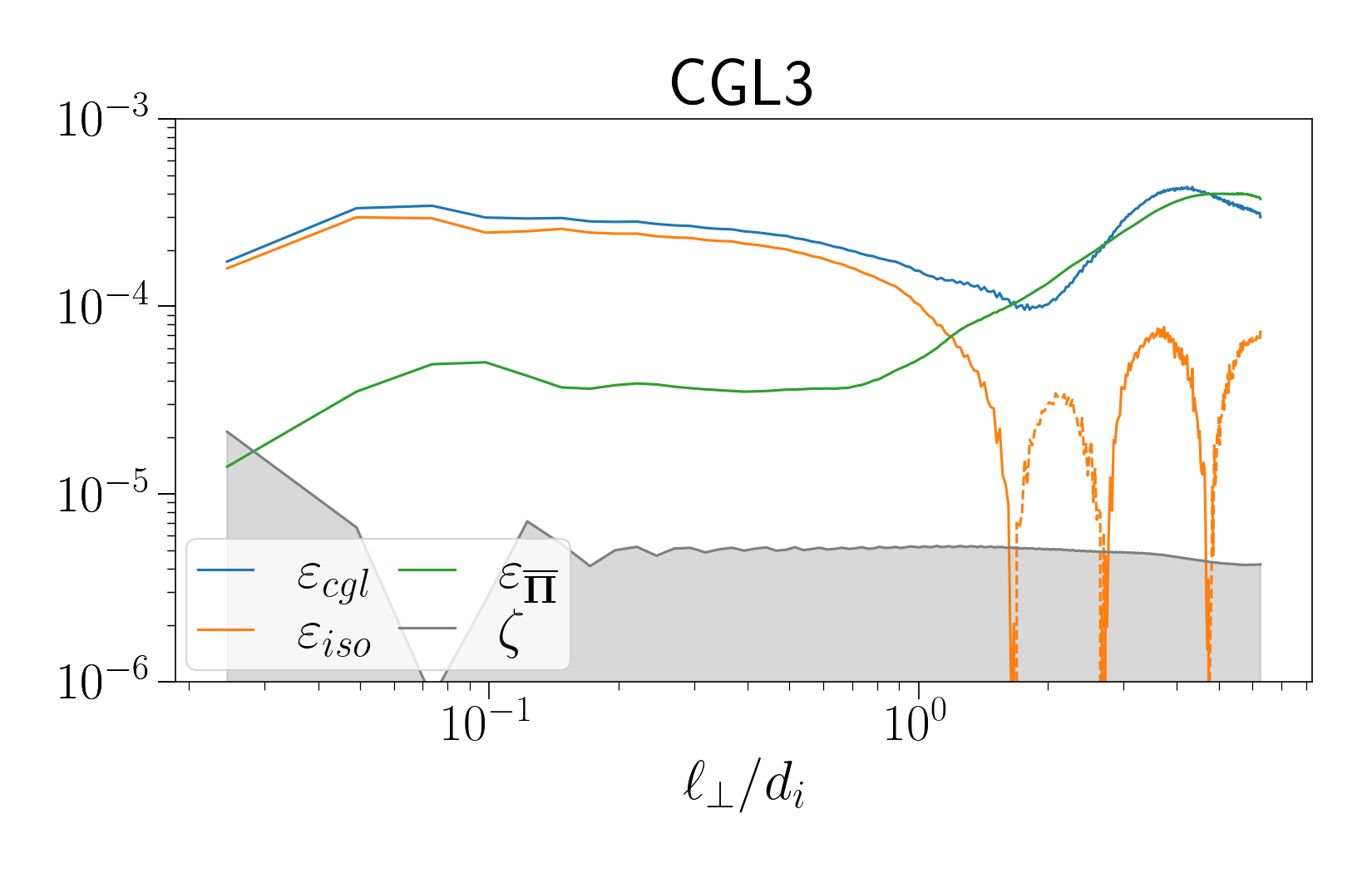}
\includegraphics[width=0.49\textwidth,trim = 1cm 2.5cm 1cm 1cm, clip]{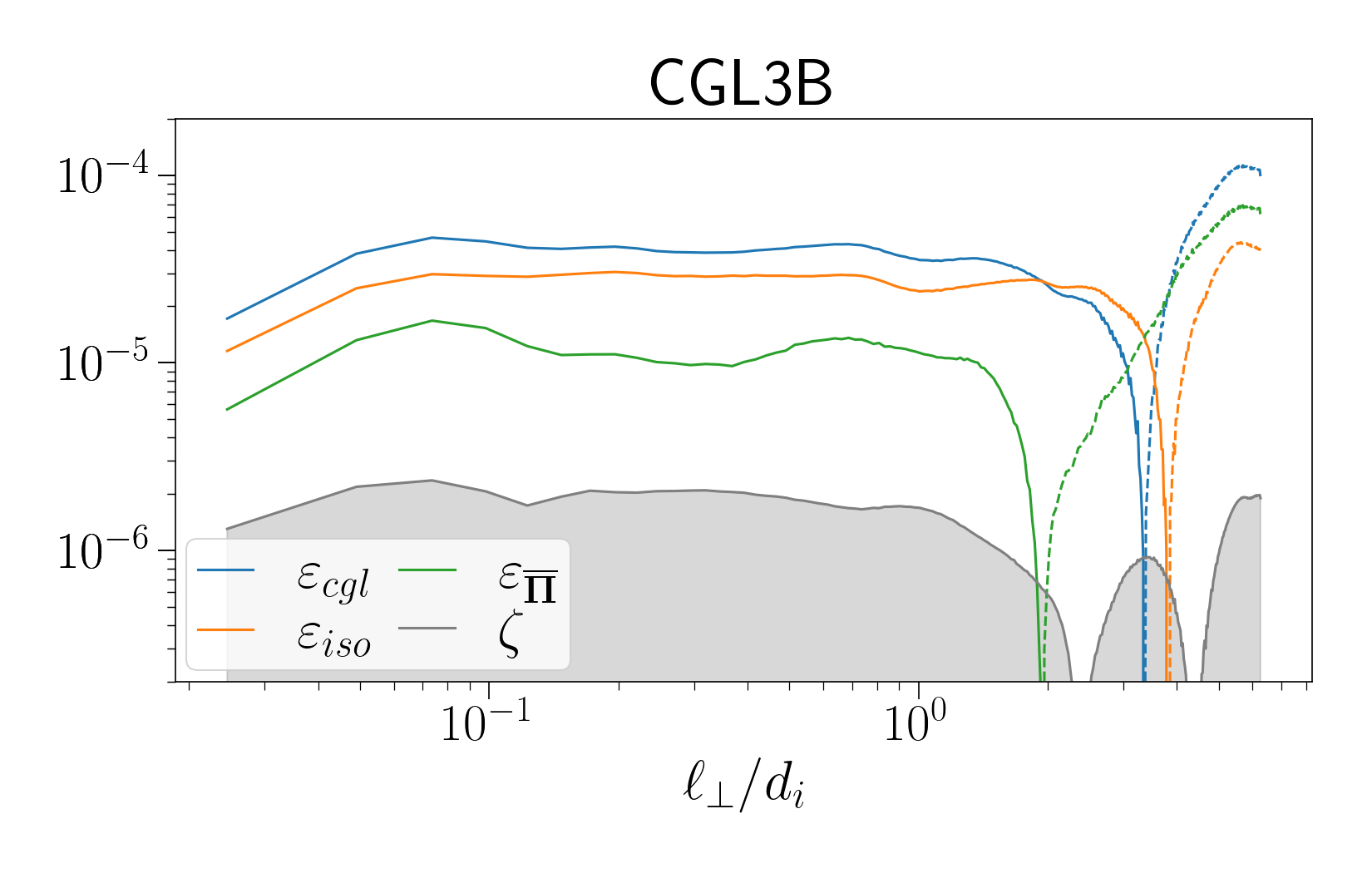}
\includegraphics[width=0.49\textwidth,trim = 1cm 1cm 1cm 1cm, clip]{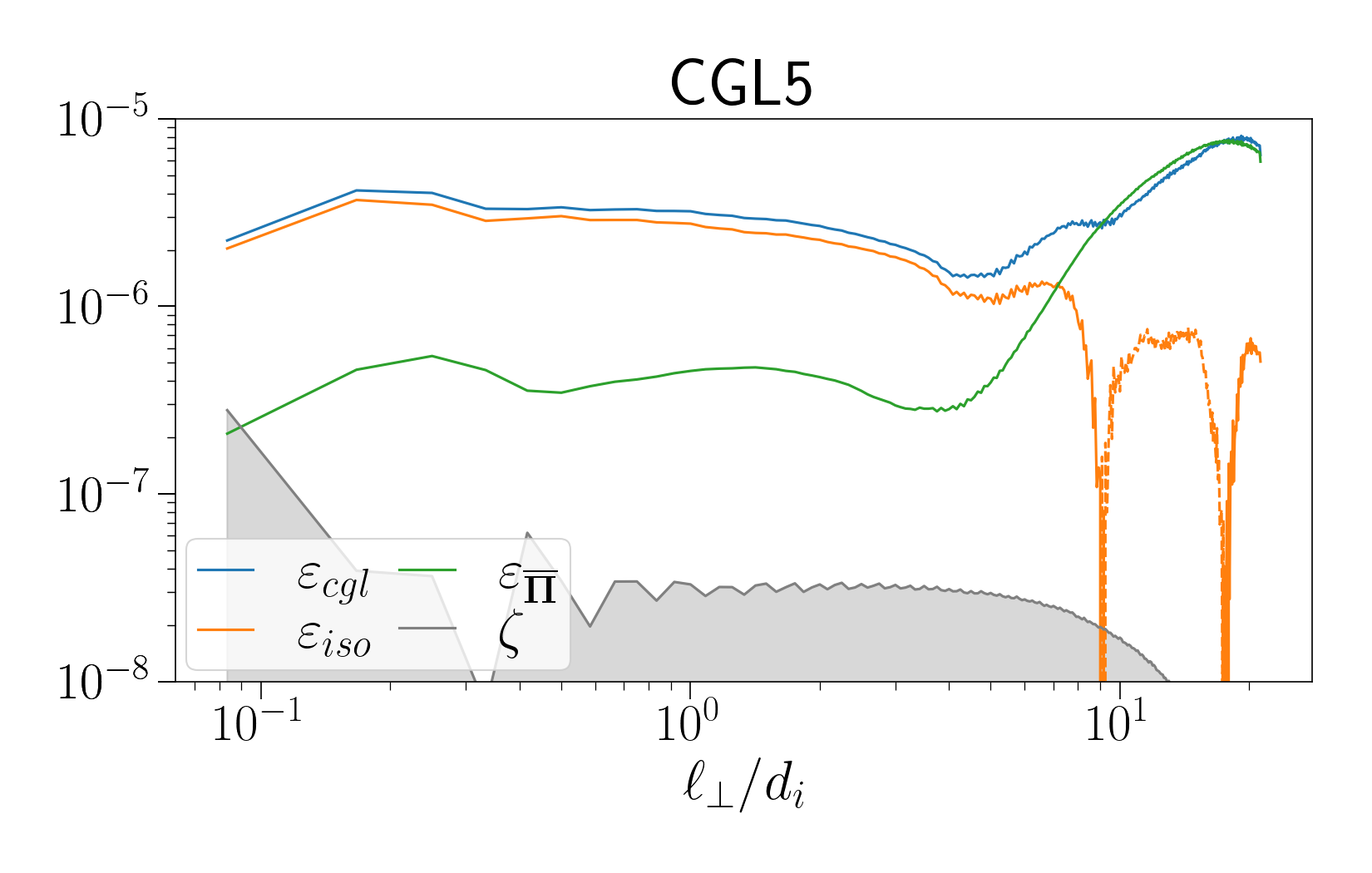}
\includegraphics[width=0.49\textwidth,trim = 1cm 1cm 1cm 1cm, clip]{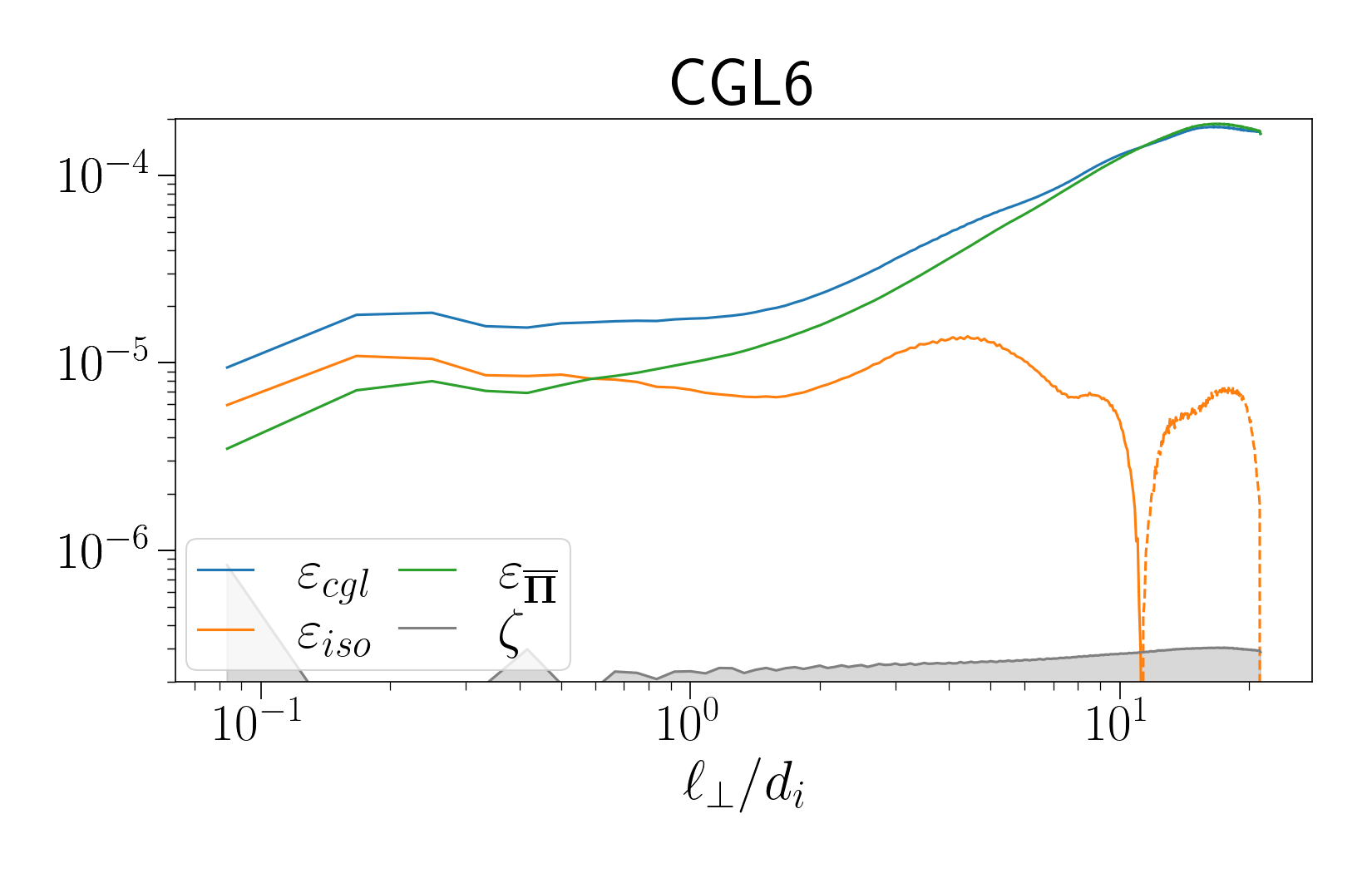}
\caption{1D perpendicular-cascade $\varepsilon_{cgl}$ (blue), $\varepsilon_{iso}$ (orange) and $\varepsilon_{\overline{\boldsymbol{\Pi}}}$ (green) for the various simulations. The gray areas hold for the error level.}
\label{fig:detail_cgl_perp}
\end{figure}

\begin{figure}
\center
\includegraphics[width=0.506\textwidth,trim = 0cm 1.2cm 1.8cm 1.7cm, clip]{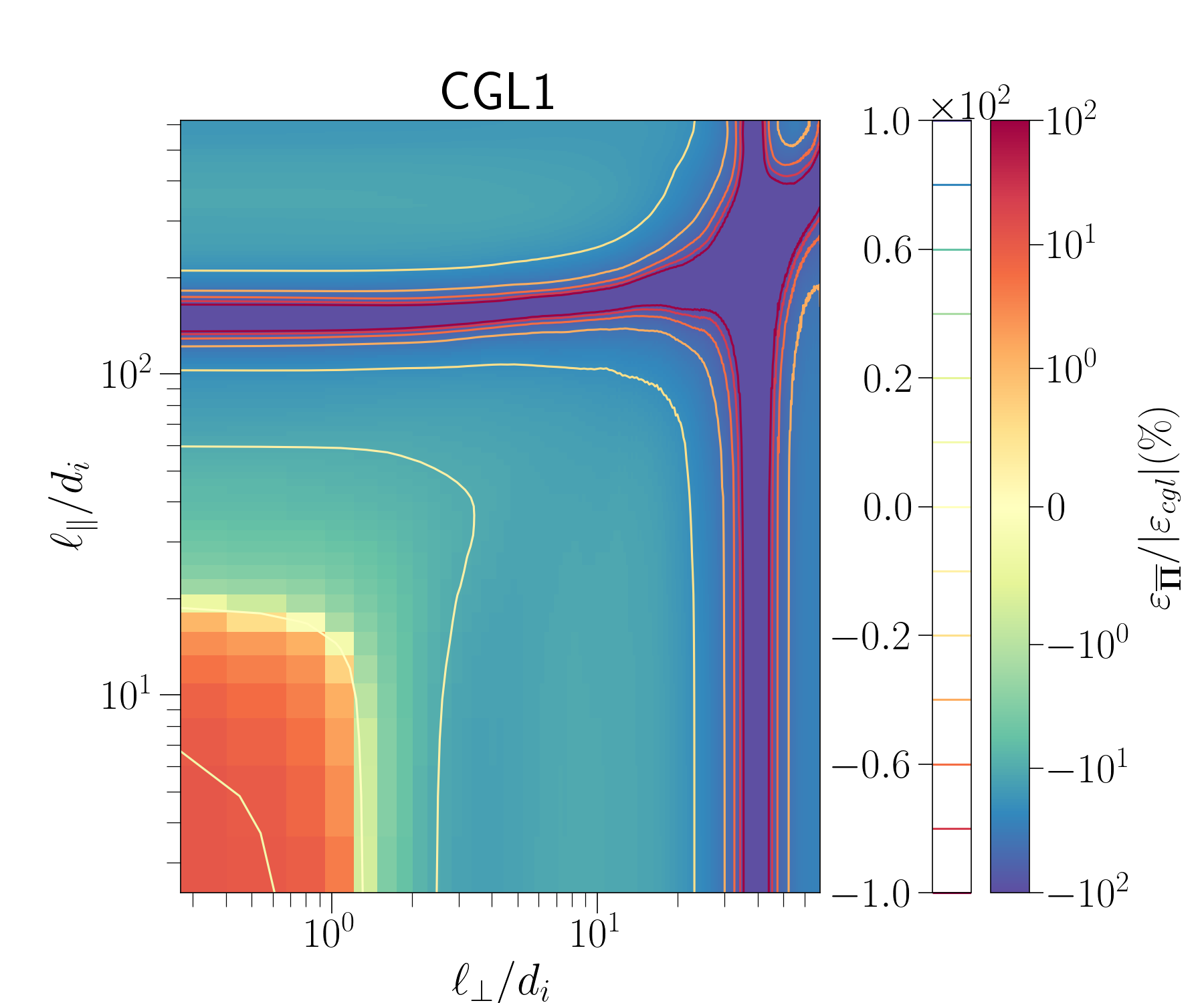}
\includegraphics[width=0.486\textwidth,trim = 2.5cm 1.2cm 0.1cm 1.7cm, clip]{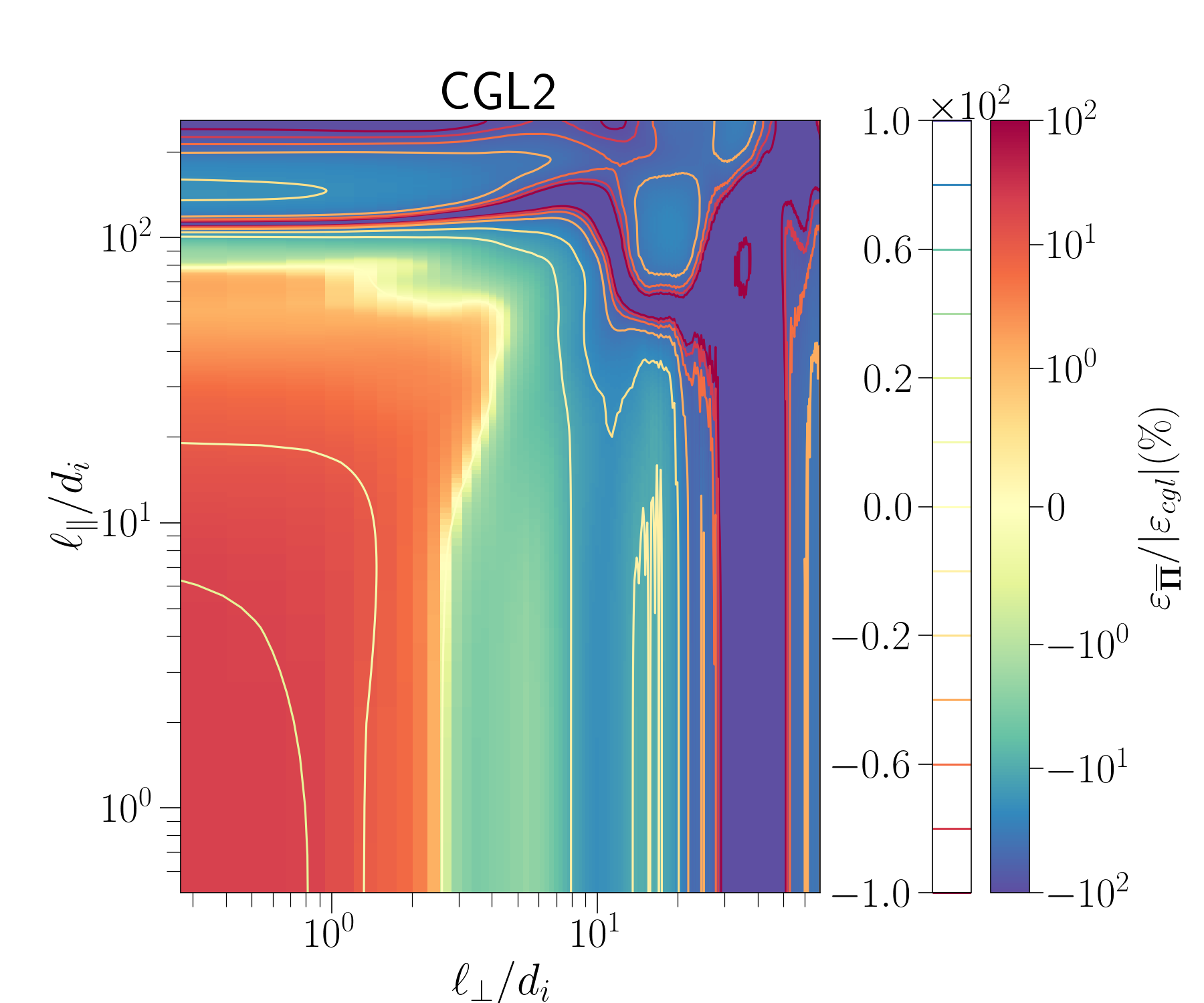}
\includegraphics[width=0.501\textwidth,trim = 0cm 1.2cm 1.8cm 1.7cm, clip]{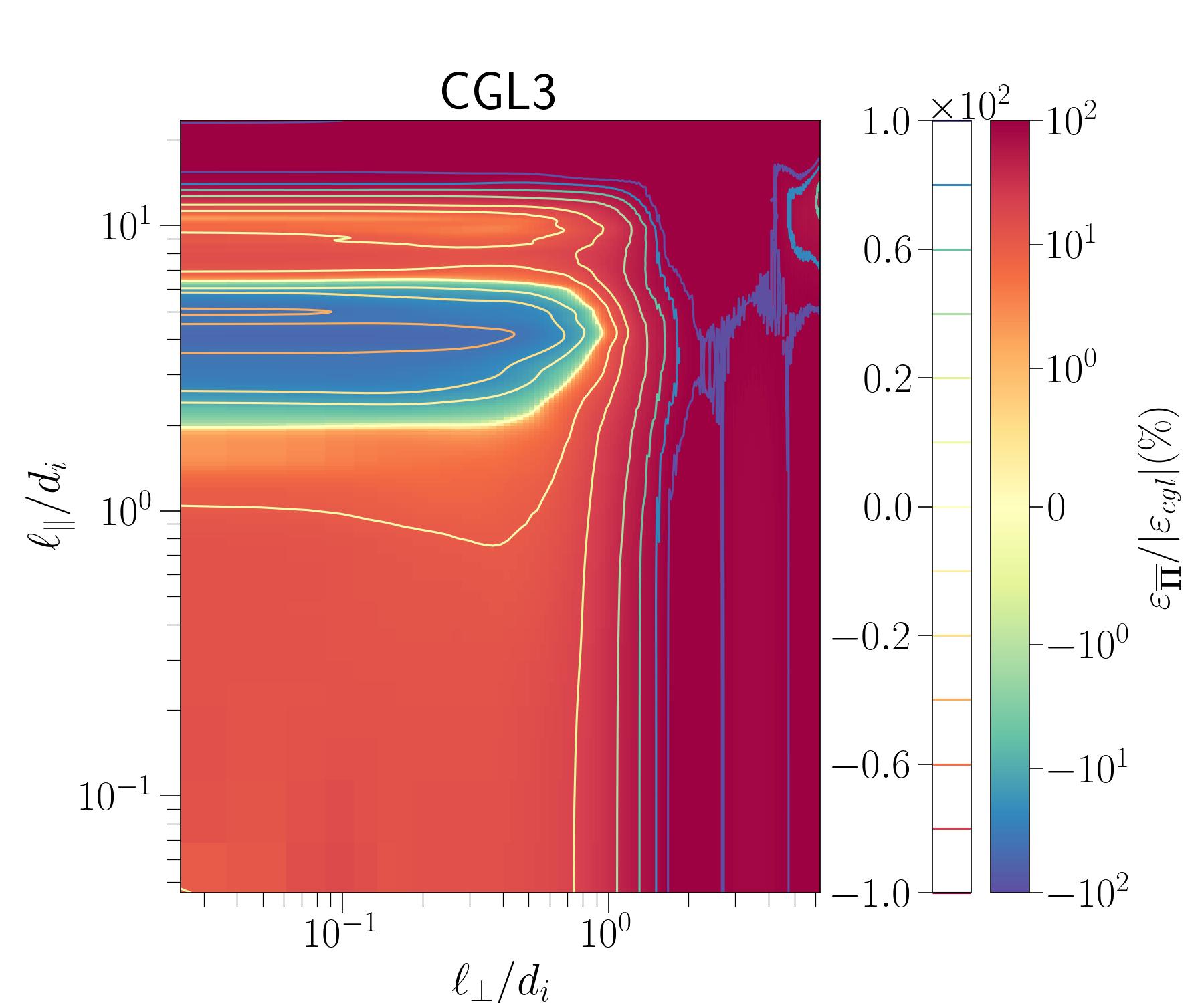}
\includegraphics[width=0.491\textwidth,trim = 2cm 1.2cm 0cm 1.7cm, clip]{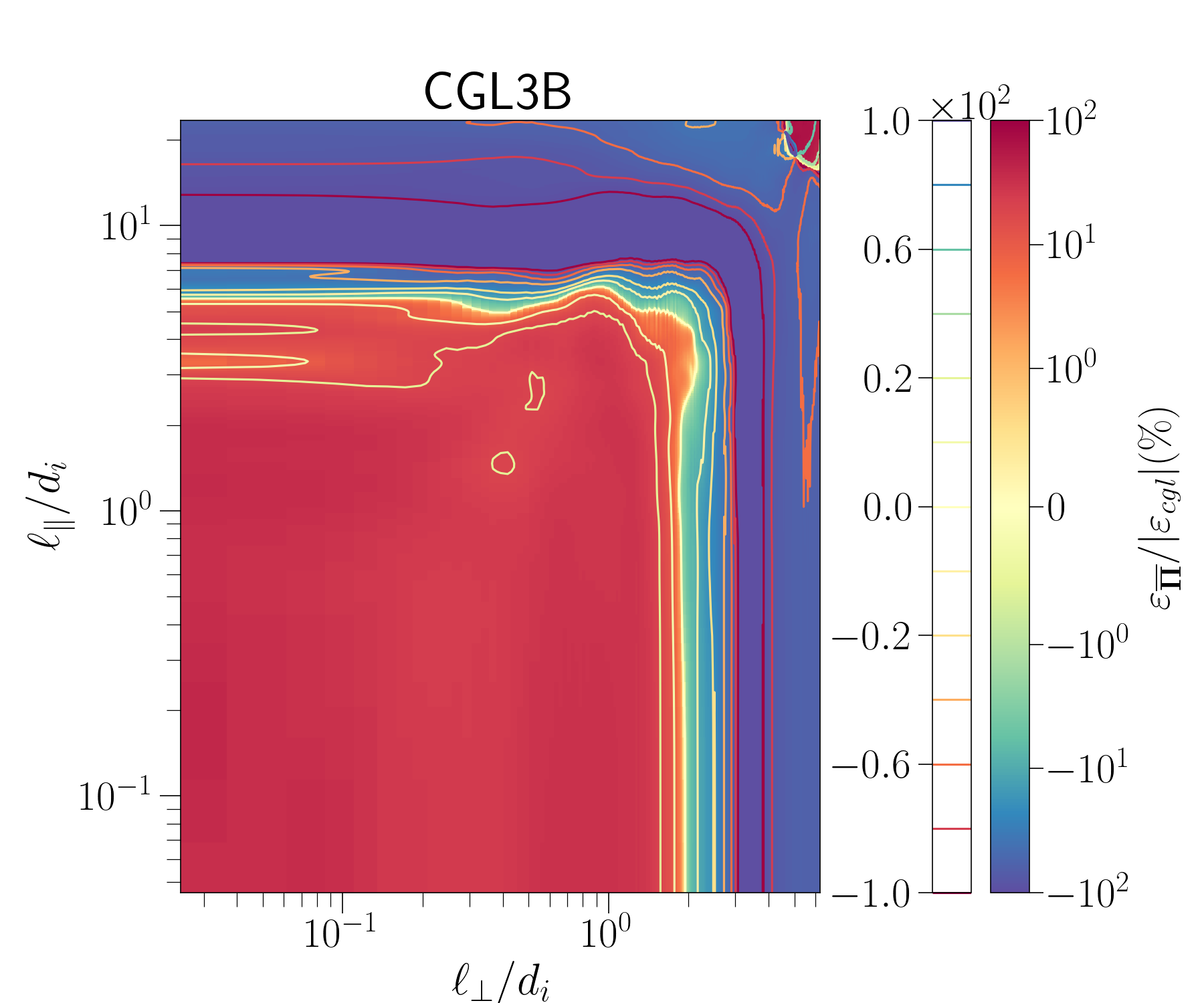}
\includegraphics[width=0.506\textwidth,trim = 0cm 0cm 1.8cm 1.7cm, clip]{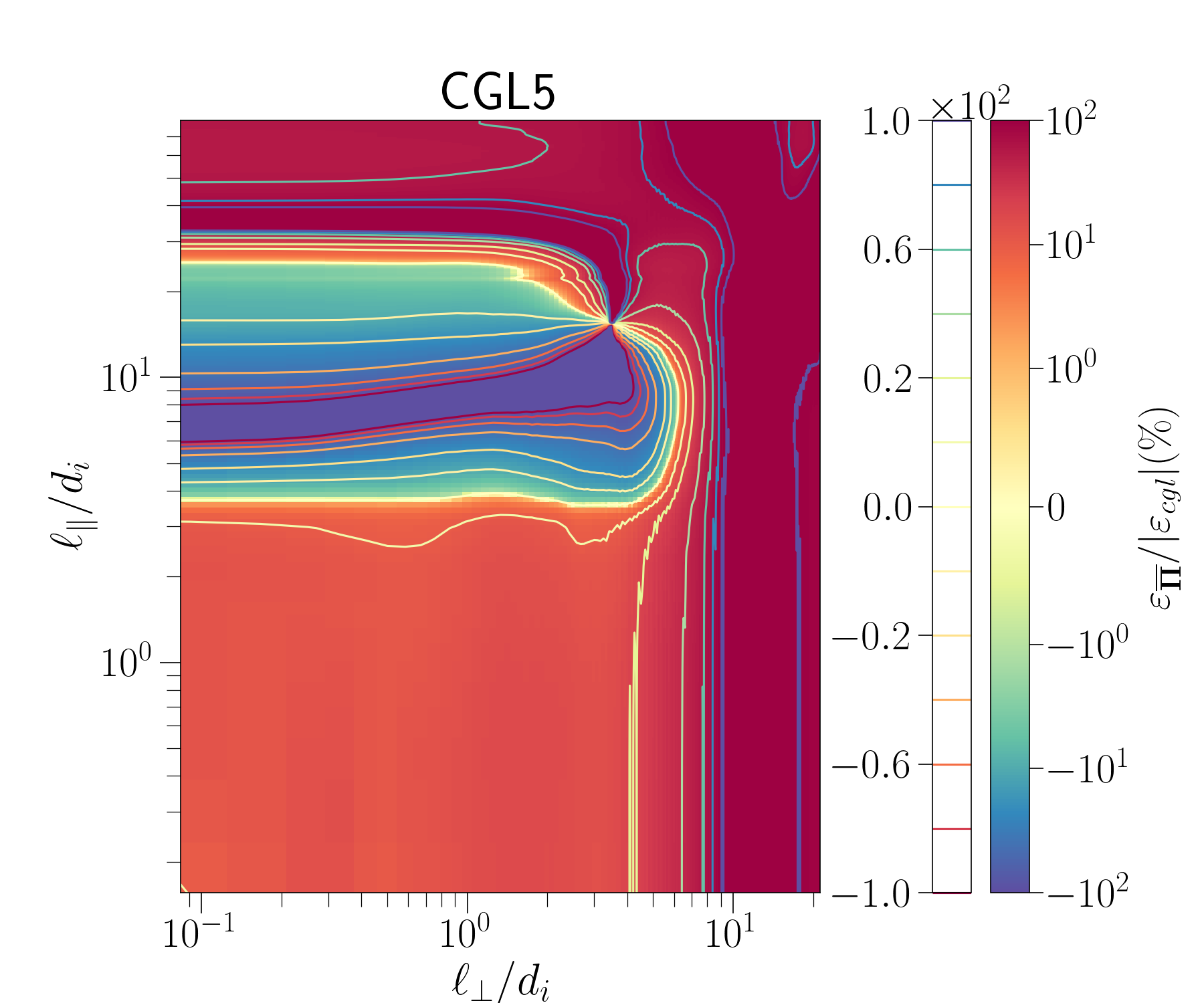}
\includegraphics[width=0.486\textwidth,trim = 2.5cm 0cm 0.1cm 1.7cm, clip]{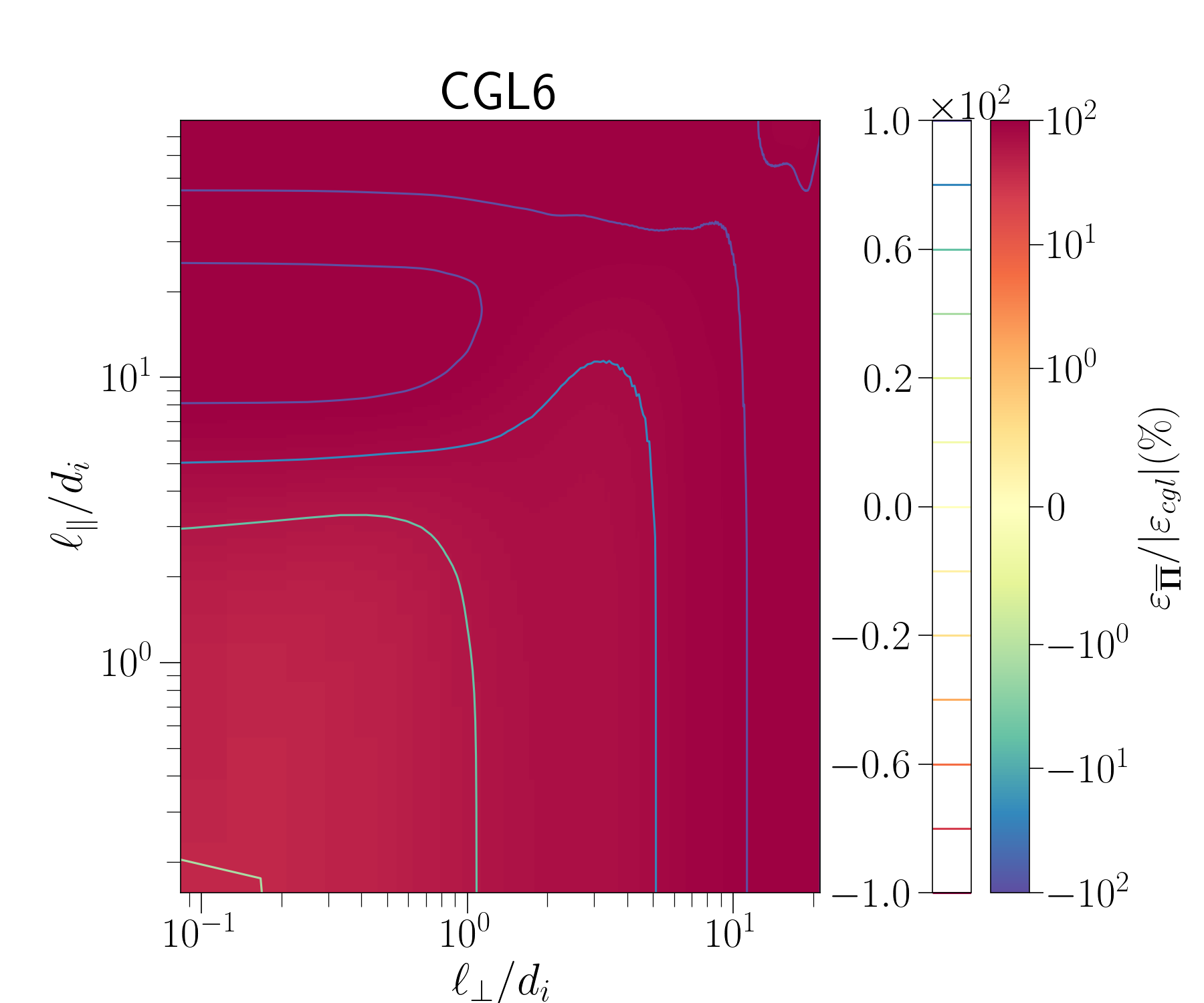}
\caption{2D map of $\varepsilon_{\overline{\boldsymbol{\Pi}}}/|\varepsilon_{cgl}|$ for the different runs. Warm color indicates a positive contribution of pressure anisotropy varying from 0 (yellow) to $100 \%$ (red), and green-blue color, a negative contribution, while the purple domain circled by a red contour line corresponds to the range where the sign of $\varepsilon_{cgl}$ fluctuates, because of the variations of $\varepsilon_{iso}$.}
\label{fig:map_pivscgl}
\end{figure}

We now analyze, in the various simulations, the partition of the total cascade rate into $\varepsilon_{\overline{\boldsymbol{\Pi}}}$, originating from the pressure anisotropy, and the isotropic contribution $\varepsilon_{iso}$. The results are shown in Fig. \ref{fig:detail_cgl_perp} as 1D perpendicular profiles in $\ell_\perp$. An estimate of the pressure-anisotropy contribution, through the signed ratio $\frac{\varepsilon_{\overline{\boldsymbol{\Pi}}}}{|\varepsilon_{cgl}|}$, is given in Fig. \ref{fig:map_pivscgl}, 
using $(\ell_\perp,\ell_\|)$ 2D maps. 

Figure \ref{fig:detail_cgl_perp} shows that the contribution of $\varepsilon_{\overline{\boldsymbol{\Pi}}}$ to the total cascade rate in the inertial range is smaller ($\varepsilon_{\overline{\boldsymbol{\Pi}}} \simeq 10 \% $ of $\varepsilon_{cgl}$) for CGL1 and CGL2 than for CGL3B ($30 \%$) and CGL6 ($50 \%$). However, at the largest scales, close to the forcing, this contribution reaches $30-60\%$ of $\varepsilon_{cgl}$ for CGL1-2-3B, and $100\%$ for CGL3-5-6. This increase is also visible in Fig. \ref{fig:map_pivscgl}, where the maps become bluer or redder at large scales. The higher level of the full cascade rate for CGL6 compared to the other runs is due to the larger initial pressure anisotropy of this run, which impacts $\varepsilon_{\overline{\boldsymbol{\Pi}}}$.
One also notices the similarities in the behavior of CGL3 and CGL5. The case of CGL3B will be discussed in Section \ref{sec:disc}. As confirmed by CGL5, for which the forcing scale lies between those of CGL1-2 and of CGL3B-3, the variations of the cascade rate occur mainly at the injection scales.

As seen in Fig. \ref{fig:detail_cgl_perp}, for CGL1, CGL2 and CGL3B, $\varepsilon_{\overline{\boldsymbol{\Pi}}}$ is positive at the smallest scales ($\ell_\perp/d_i\lesssim 1$) and negative at larger ones. The change of sign occurs in the MHD range, close to $\ell_{\perp}=d_i$ in the perpendicular direction, and close to the forcing scales in the parallel direction (see Fig. \ref{fig:map_pivscgl}). This behavior underlines the role of the forcing in the generation of sign fluctuations for $\varepsilon_{\overline{\boldsymbol{\Pi}}}$ (and possibly for other components of the cascade rate), rather than a physical effect occurring at the specific scales where the sign change is observed. These changes of sign induce a shallow decrease of the total cascade rate and limit the extension of the inertial range close to the forcing scale. While no change of sign of $\varepsilon_{\overline{\boldsymbol{\Pi}}}$ is visible for the perpendicular cascade of CGL3 and CGL5, the cascade rate behaves differently in the parallel direction for these simulations. CGL6 shows no sign change in either direction. For all the simulations, $\varepsilon_{iso}$ and $\varepsilon_{\overline{\boldsymbol{\Pi}}}$ are significantly larger than the uncertainty level $\zeta$, indicating a negligible effect of the numerical error on the total cascade rate. 

The above results highlight the contribution of pressure anisotropy to the cascade rate. The detailed analysis of $\varepsilon_{iso}$ and $\varepsilon_{\overline{\boldsymbol{\Pi}}}$ is performed in the following subsections. 

\subsection{Contribution to $\varepsilon_{iso}$ induced by density fluctuations} \label{subsec:res_iso}

Here, we compare $\varepsilon_{iso}$ that collects all the terms independent of the pressure anisotropy in $\varepsilon_{cgl}$, as obtained in the present simulations, to the predictions of the incompressible cascade rate computed in 
\citet{ferrand_fluid_2021}, and with other results based on isotropic compressible models \citep{andres_energy_2018}. 
\begin{figure}
 \center
 \includegraphics[width=0.49\textwidth,trim = 0.5cm 3cm 0cm 5cm, clip]{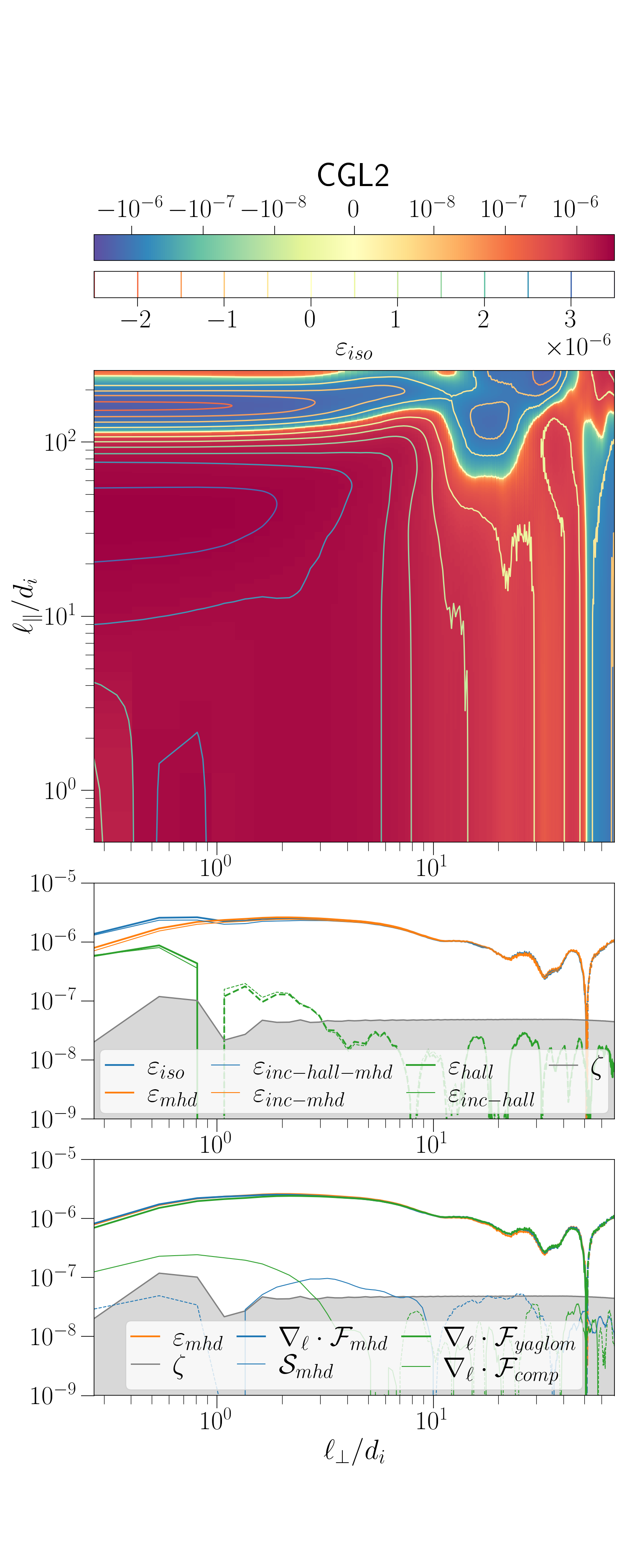} 
 \includegraphics[width=0.475\textwidth,trim = 1.2cm 3cm 0cm 5cm, clip]{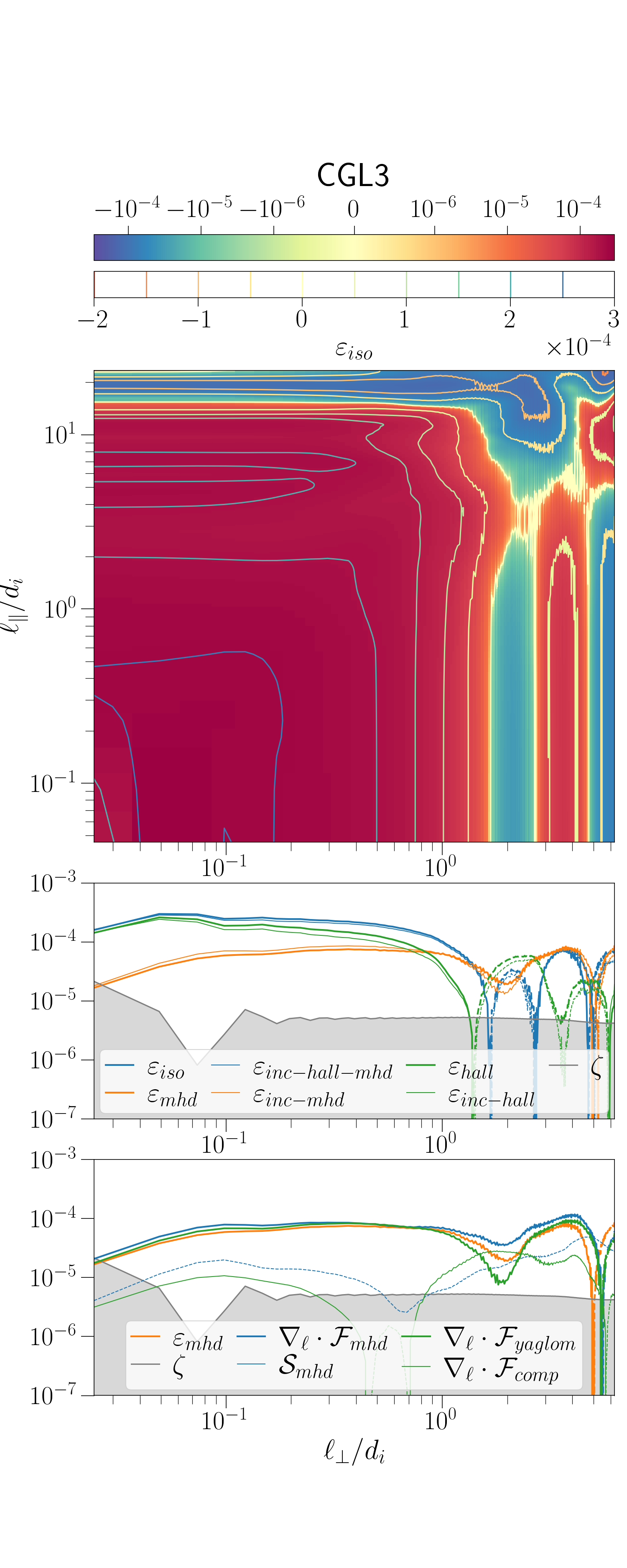}
 \caption{Left: CGL2. Right: CGL3. Top: 2D map (with same graphic conventions as in Fig. \ref{fig:epsCGL}) of $\varepsilon_{iso}$ in the $(\ell_{\perp},\ell_{\parallel})$ plane. Middle and bottom: 1D perpendicular profiles of the various contributions to $\varepsilon_{iso}$. Middle: comparison of the MHD ($\varepsilon_{mhd}$, thick orange) and Hall ($\varepsilon_{hall}$, thick green) contributions to $\varepsilon_{iso}$ (thick blue) with their incompressible counterparts (thin curves). Bottom: components of $\varepsilon_{mhd}$ (orange) i.e. the flux part (thick blue) and source part (thin blue), and detail of the flux part i.e. Yaglom part (thick green) and purely compressible part (thin green).}
 \label{fig:decomp_iso}
\end{figure}

Figure \ref{fig:decomp_iso} shows the $(\ell_{\perp},\ell_{\parallel})$ 2D map of $\varepsilon_{iso}$, obtained from CGL2 (left) and CGL3 (right). An inertial range is visible between $\ell_{\perp} \in [0.5, 7]$ and $\ell_{\parallel} \in [5, 90]$ for CGL2, and between $\ell_{\perp} \in [0.05, 1]$ and $\ell_{\parallel} \in [0.2, 10]$ for CGL3. At the smallest scales, $\varepsilon_{iso}$ decreases due to dissipation, and varies in sign and amplitude at the largest scales where the forcing acts. As expected, the inertial range (in the sense of a constant cascade rate) of CGL2 is evidenced in the MHD scales, while for CGL3 it is limited to the Hall scales. CGL1 (not shown) exhibits a broader inertial range in $\ell_\perp$ than CGL2, probably due to a more efficient generation of Alfvenic turbulence by a driving acting in a more quasi-perpendicular direction ($\theta_i=83^\circ$ for CGL1 and $75^\circ$ for CGL2).
 
These observations also hold for the 1D perpendicular cascade rate displayed in the middle graphs of Fig. \ref{fig:decomp_iso} (thick blue line) for each simulation. The profile of $\varepsilon_{iso}$ is close to that of the incompressible Hall-MHD contribution ($\varepsilon_{inc-hall-mhd}$, thin blue line), obtained with the third-order law used by \citet{ferrand_exact_2019}. The MHD ($\varepsilon_{mhd}$, orange) and Hall ($\varepsilon_{hall}$, green) components are also close to their incompressible counterparts ($\varepsilon_{inc-mhd}$ and $\varepsilon_{inc-hall}$, thin lines): the difference between compressible and incompressible models vary between $1\%$ and $10\%$, and is slightly more important for CGL3. Such observations are expected since the present simulations are only weakly compressible (for CGL2: $\frac{std(\rho)}{\left< \rho \right>} = 2\%$ for CGL3: $\frac{std(\rho)}{\left< \rho \right>} = 8\%$). The MHD and Hall contributions behave as expected: at scales $\ell_{\perp}\sim d_i$, the Hall contribution increases and overcomes the decreasing MHD one, thus ensuring a constant total cascade rate. 
 
The various components of $\varepsilon_{mhd}$ are displayed in the bottom panels of Fig. \ref{fig:decomp_iso} for each simulation, with $\varepsilon_{mhd}$ decomposed as $\varepsilon_{mhd} = \bnabla_{\boldsymbol{\ell}} \cdot \boldsymbol{\mathcal{F}}_{mhd} + \mathcal{S}_{mhd}$ where
 \begin{equation}
 \boldsymbol{\mathcal{F}}_{mhd} = \boldsymbol{\mathcal{F}}_{yaglom} + \boldsymbol{\mathcal{F}}_{comp} 
 \end{equation}
 with 
 \begin{eqnarray}
 && \left\{
 \begin{split} \boldsymbol{\mathcal{F}}_{yaglom} &= -\frac{1}{4} \langle \left(\delta \left(\rho \boldsymbol{v}\right) \cdot \delta \boldsymbol{v} + \delta \left(\rho \boldsymbol{v_A}\right) \cdot \delta \boldsymbol{v_A} \right) \delta \boldsymbol{v} 
 - \left( \delta \left(\rho \boldsymbol{v_A}\right) \cdot \delta \boldsymbol{v} + \delta \left(\rho \boldsymbol{v}\right) \cdot \delta \boldsymbol{v_A} \right) \delta \boldsymbol{v_A} \rangle \, ,\\
 \boldsymbol{\mathcal{F}}_{comp} &= -\frac{1}{4} \langle \delta \rho \left(2 \delta u - \delta \left(\frac{p+p_m}{\rho}\right)\right) \delta \boldsymbol{v} \rangle \, \end{split}
 \right. \nonumber 
 \end{eqnarray}
 and
 \begin{equation}
 \mathcal{S}_{mhd} = \left(\mathcal{S}(\boldsymbol{\ell}) + \mathcal{S}(-\boldsymbol{\ell}) \right) 
 \end{equation}
 with 
 \begin{eqnarray}
 &&\left\{
 \begin{split} \mathcal{S}(\boldsymbol{\ell}) & = \langle \left(\rho \boldsymbol{v} \cdot \delta \boldsymbol{v} +\frac{1}{2} \rho \boldsymbol{v_A} \cdot \delta \boldsymbol{v_A} - \frac{1}{2}\boldsymbol{v_A} \cdot \delta \left(\rho \boldsymbol{v_A}\right) + 2 \rho \delta u\right)\bnabla' \bcdot \boldsymbol{v'} \rangle \\
 &+ \langle \left( - 2 \rho \boldsymbol{v} \cdot \delta \boldsymbol{v_A} - \rho \boldsymbol{v_A} \cdot \delta \boldsymbol{v} + \delta (\rho \boldsymbol{v}) \cdot \boldsymbol{v_A} \right) \bnabla' \bcdot \boldsymbol{v'_A} \rangle \\
 &+ \langle - 2\rho \delta \left(\frac{p}{\rho}\right) : \bnabla' \boldsymbol{v'} \rangle+ \langle\left(\delta \rho \frac{p+p_m}{\rho} \cdot \boldsymbol{v} - \rho \delta \left(\frac{p+p_m}{\rho}\right) \cdot \boldsymbol{v} \right)\cdot \frac{\bnabla' \rho' }{\rho'} \rangle, \end{split}
 \right. \nonumber
 \end{eqnarray}
The quantity $\bnabla \cdot \boldsymbol{\mathcal{F}}_{yaglom}$ is the part of $\varepsilon_{mhd}$ that survives when formally taking $\rho=\rho_0=1$, typical of an incompressible regime, yielding $\varepsilon_{inc-mhd}$, while $\bnabla \cdot \boldsymbol{\mathcal{F}}_{comp}$ will vanish in this limit. Note that $\bnabla \cdot \boldsymbol{\mathcal{F}}_{comp}$ and $\mathcal{S}_{mhd}$, referred to as hybrid and source terms in \citet{andres_energy_2018} and \citet{simon_general_2021}, are expressed here using the structure functions. Similarly to the results reported in isothermal turbulence simulations \citep{andres_energy_2018}, their contribution to $\varepsilon_{mhd}$ remains negligible ($1\%$ to $10\%$ at the smallest scales), an effect possibly originating from the weak compressibility of the present simulations.

Similar behavior of $\varepsilon_{iso}$ is obtained for the other simulations, in agreement with previous results published in the literature, thus validating the contribution to $\varepsilon_{cgl}$ that is independent of the pressure anisotropy.

\subsection{Detailed analysis of $\varepsilon_{\overline{\boldsymbol{\Pi}}}$}\label{subsec:res_an}

All the simulations display similar results about the contributions of the different terms entering $\varepsilon_{\overline{\boldsymbol{\Pi}}}$. Since CGL6 is the run that shows the strongest effect of $\varepsilon_{\overline{\boldsymbol{\Pi}}}$, we concentrate here on this run and on CGL5 that is initialized with ${a_p}_0=1$, for comparison. The different terms contributing to $\varepsilon_{\overline{\boldsymbol{\Pi}}}$ are given in Eq.\eqref{eq:dtail_an}. 

\begin{figure}
\center
\includegraphics[width=0.49\textwidth,trim = 1cm 1cm 1cm 1cm, clip]{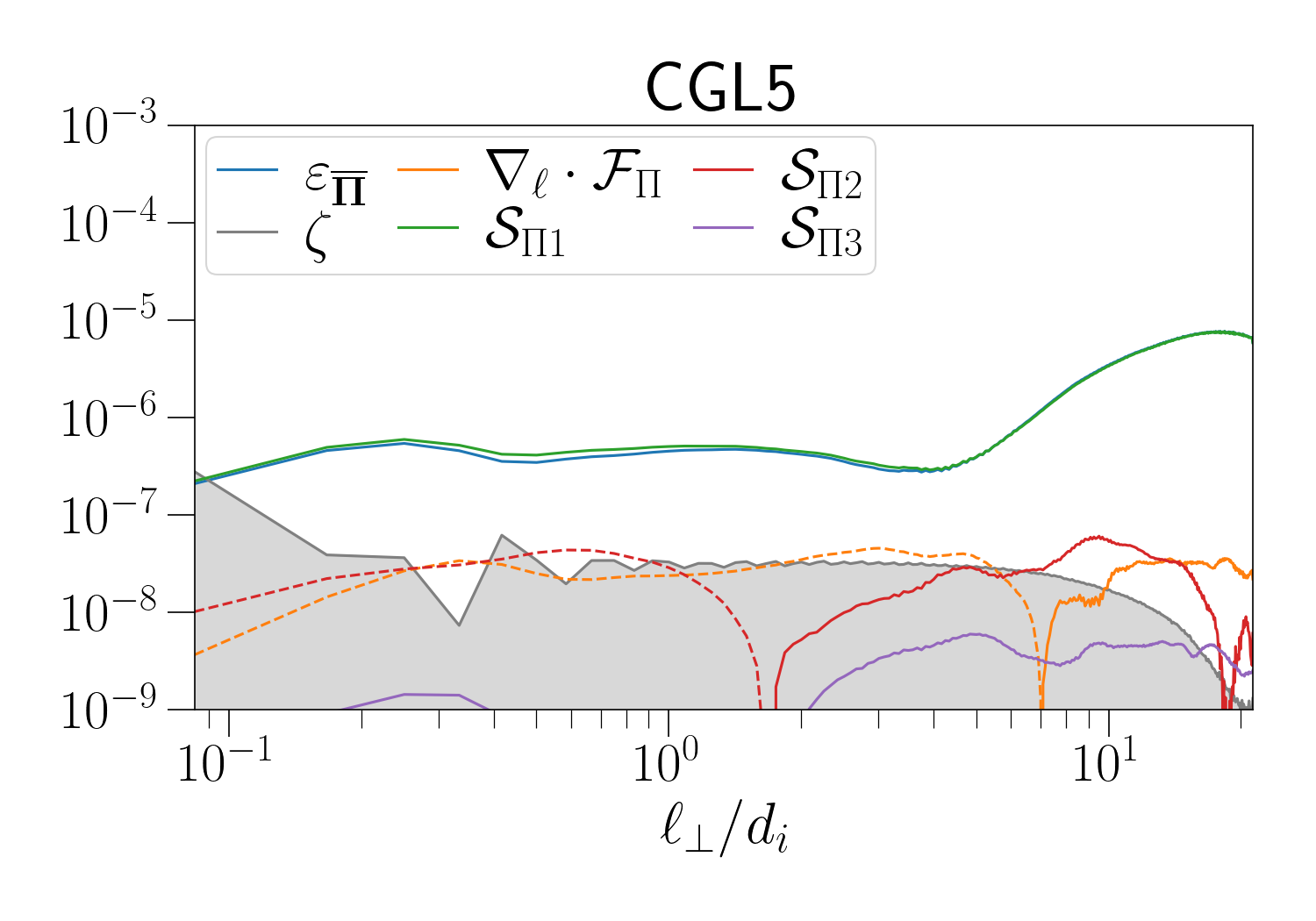}
\includegraphics[width=0.49\textwidth,trim = 1cm 1cm 1cm 1cm, clip]{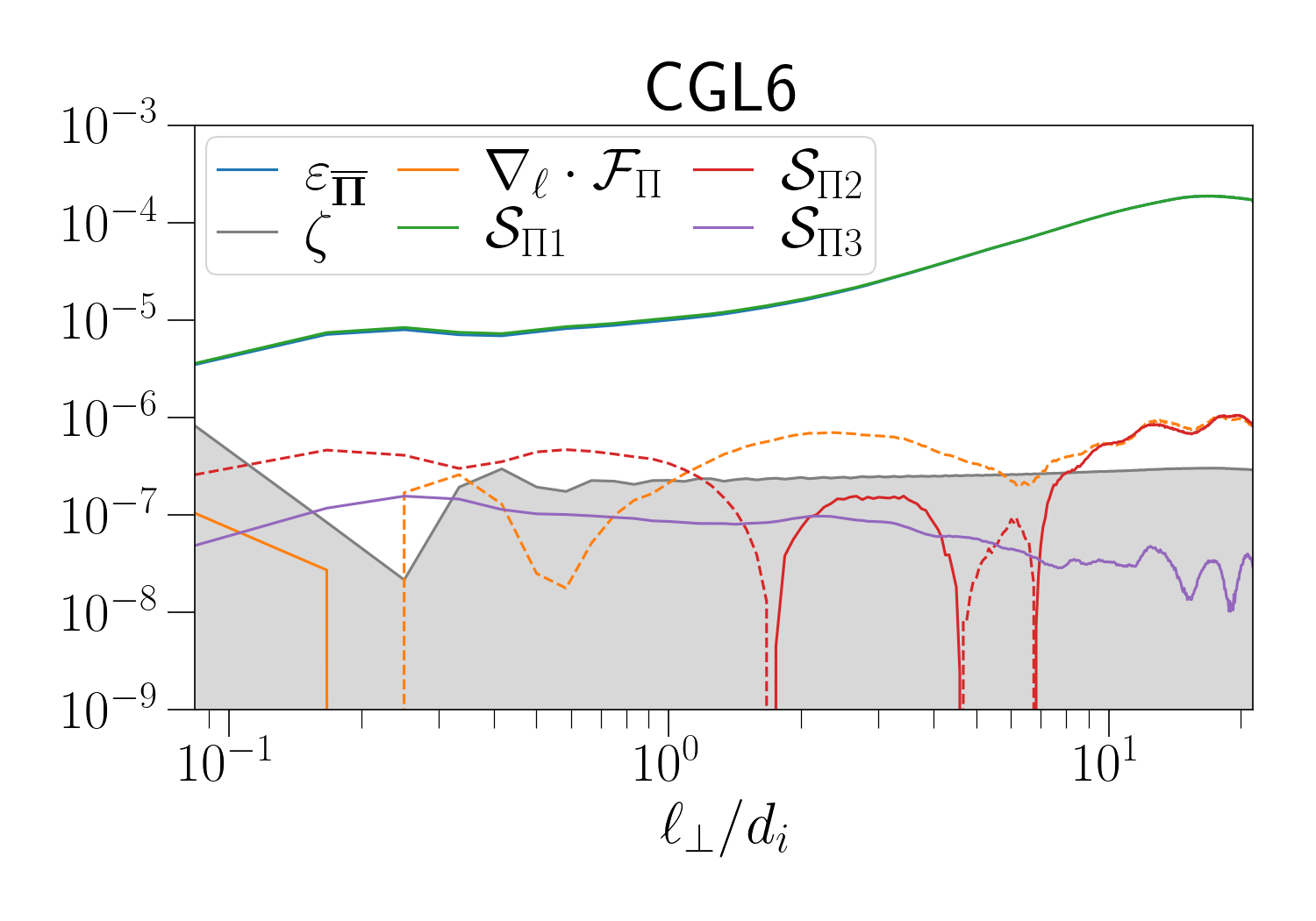}
\caption{1D perpendicular profile of the different contributions to $\varepsilon_{\overline{\boldsymbol{\Pi}}}$ (blue) in run CGL5 and CGL6. The source term that does not explicitly depend on the density fluctuations is plotted in green. The gray area reflects the uncertainty level $\zeta$. Plain lines are for positive values and dashed lines are for negative ones.} 
\label{fig:detail_pi}
\end{figure}

Figure \ref{fig:detail_pi} shows that the variations in amplitude and sign of $\varepsilon_{\overline{\boldsymbol{\Pi}}}$ (blue curve) is dominated by $\mathcal{S}_{\Pi 1}$ (green) and is only slightly impacted by the other terms whose contribution is less than $10 \%$ of $\varepsilon_{\overline{\boldsymbol{\Pi}}}$. We recall that $\mathcal{S}_{\Pi 1}$ is the term that does not explicitly depend on the density fluctuations, and thus survives in the incompressibility limit. Its impact is thus expected to be dominant in the present quasi-incompressible simulations. In order of importance, $\bnabla \cdot \boldsymbol{\mathcal{F}}_{\Pi}$ (orange) and $\mathcal{S}_{\Pi 2}$ (red) come next. These terms depend on the pressure fluctuation, and their relative importance depends on the runs: $|\bnabla \cdot \boldsymbol{\mathcal{F}}_{\Pi}|< |\mathcal{S}_{\Pi 2}|$ for CGL1 and CGL2, $|\bnabla \cdot \boldsymbol{\mathcal{F}}_{\Pi}|> |\mathcal{S}_{\Pi 2}|$ for CGL3 and CGL3B, and $|\bnabla \cdot \boldsymbol{\mathcal{F}}_{\Pi}|\sim | \mathcal{S}_{\Pi 2}|$ for CGL5 and CGL6. The last contribution $\mathcal{S}_{\Pi 3}$ (purple) that falls below the uncertainty level $\zeta$ nearly at all scales is the only source term that depends on the density fluctuations. 

\begin{figure}
\center
\includegraphics[width=0.49\textwidth,trim = 1cm 2.5cm 1cm 1cm, clip]{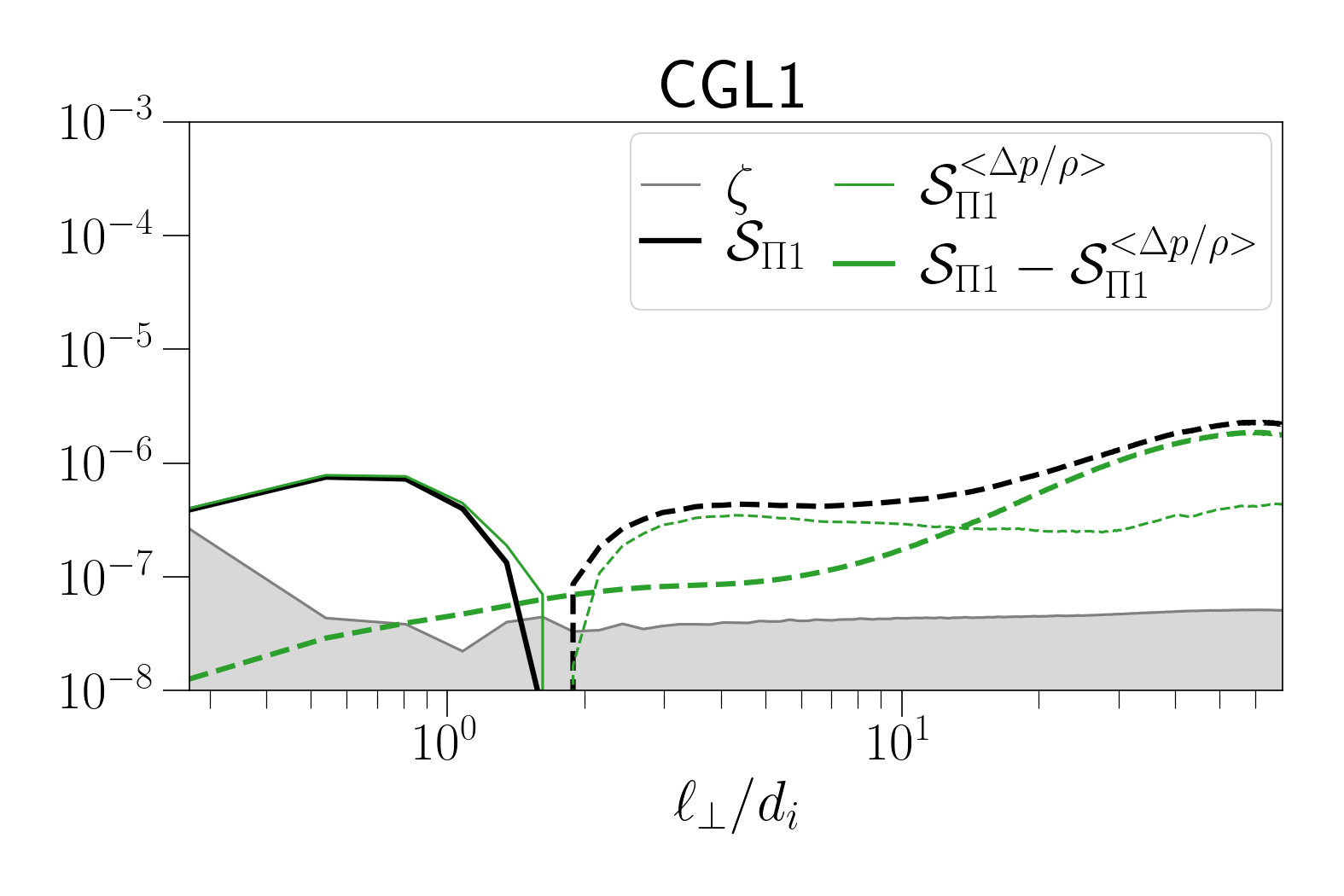}
\includegraphics[width=0.49\textwidth,trim = 1cm 2.5cm 1cm 1cm, clip]{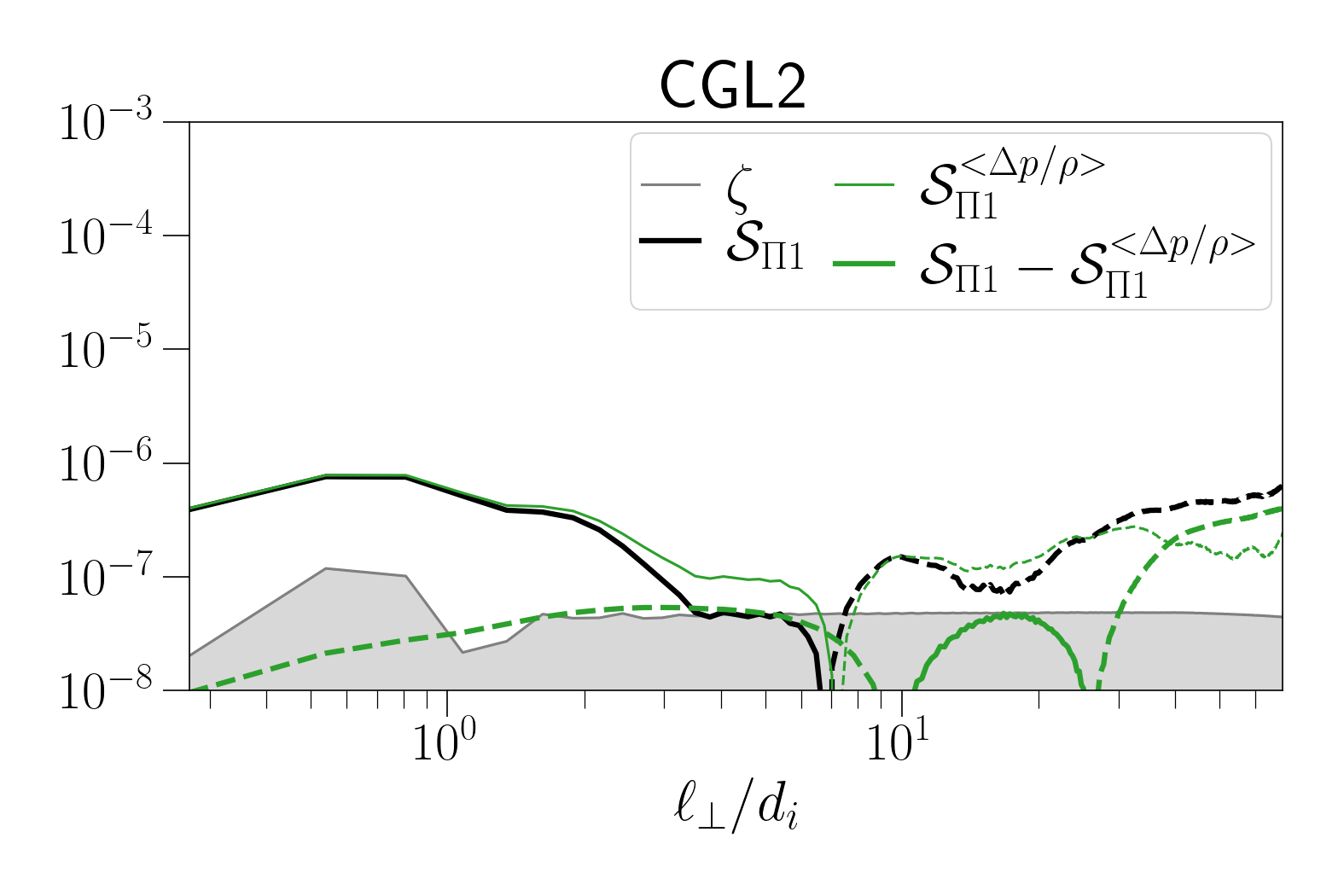}
\includegraphics[width=0.49\textwidth,trim = 1cm 2.5cm 1cm 1cm, clip]{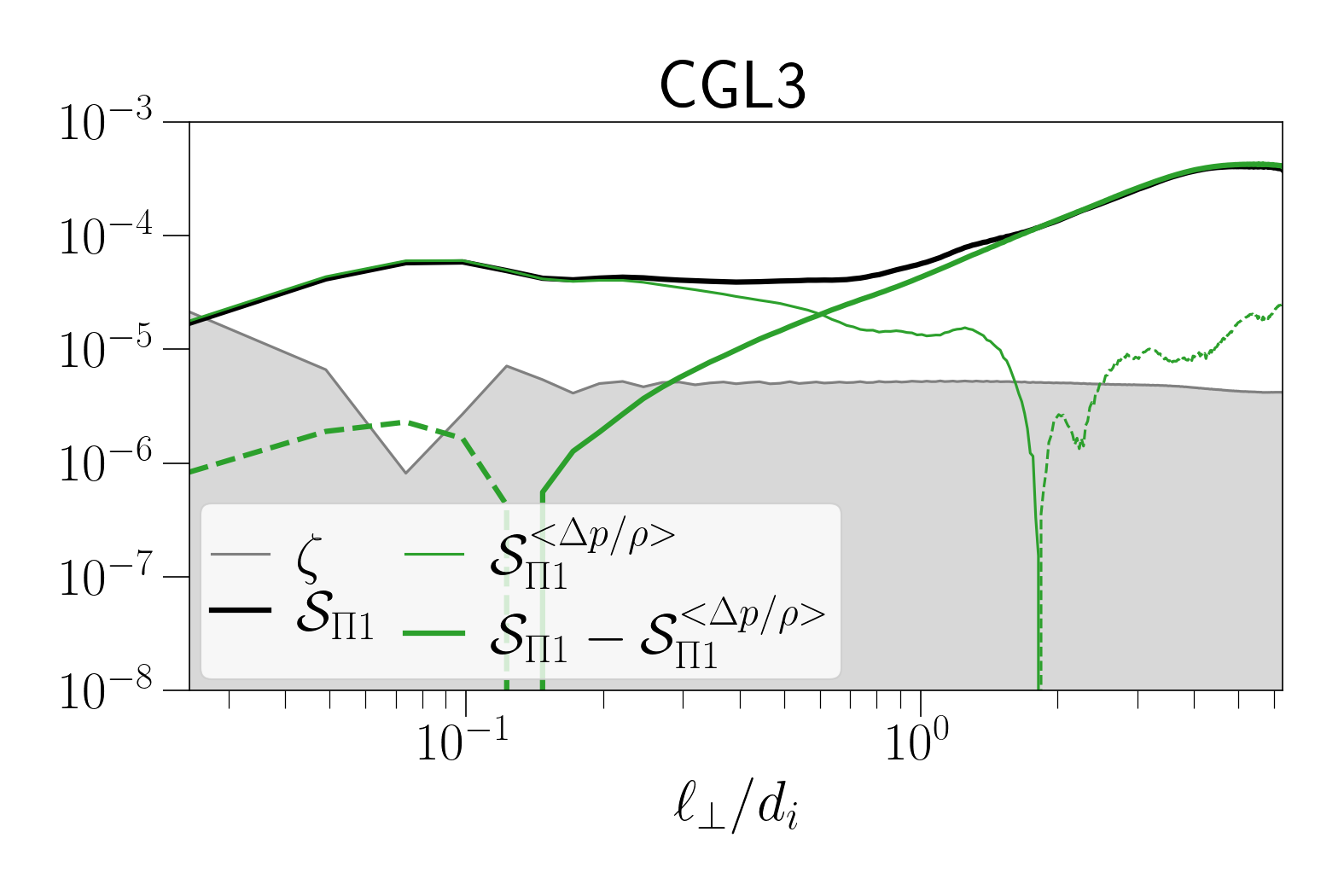}
\includegraphics[width=0.49\textwidth,trim = 1cm 2.5cm 1cm 1cm, clip]{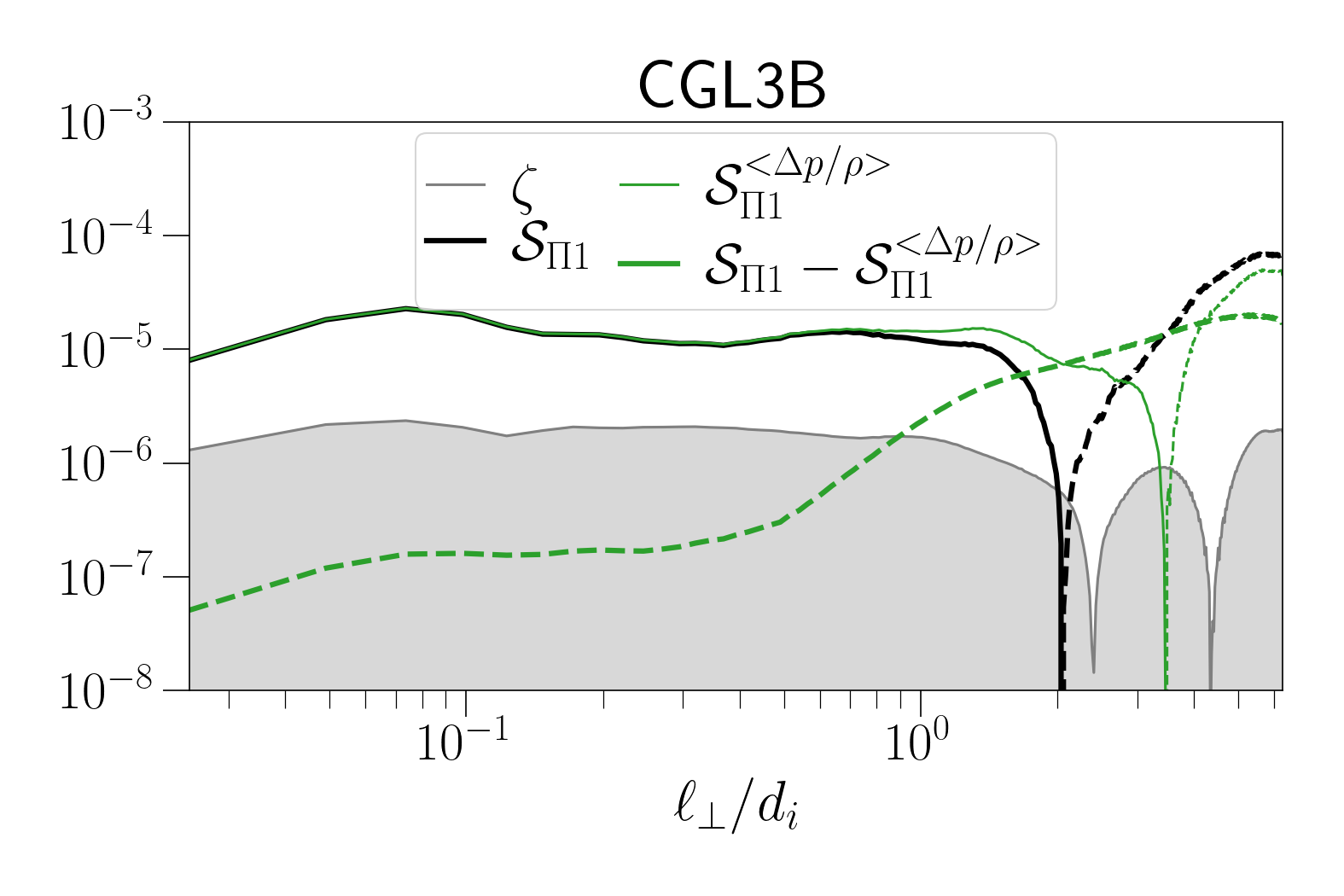}
\includegraphics[width=0.49\textwidth,trim = 1cm 1cm 1cm 1cm, clip]{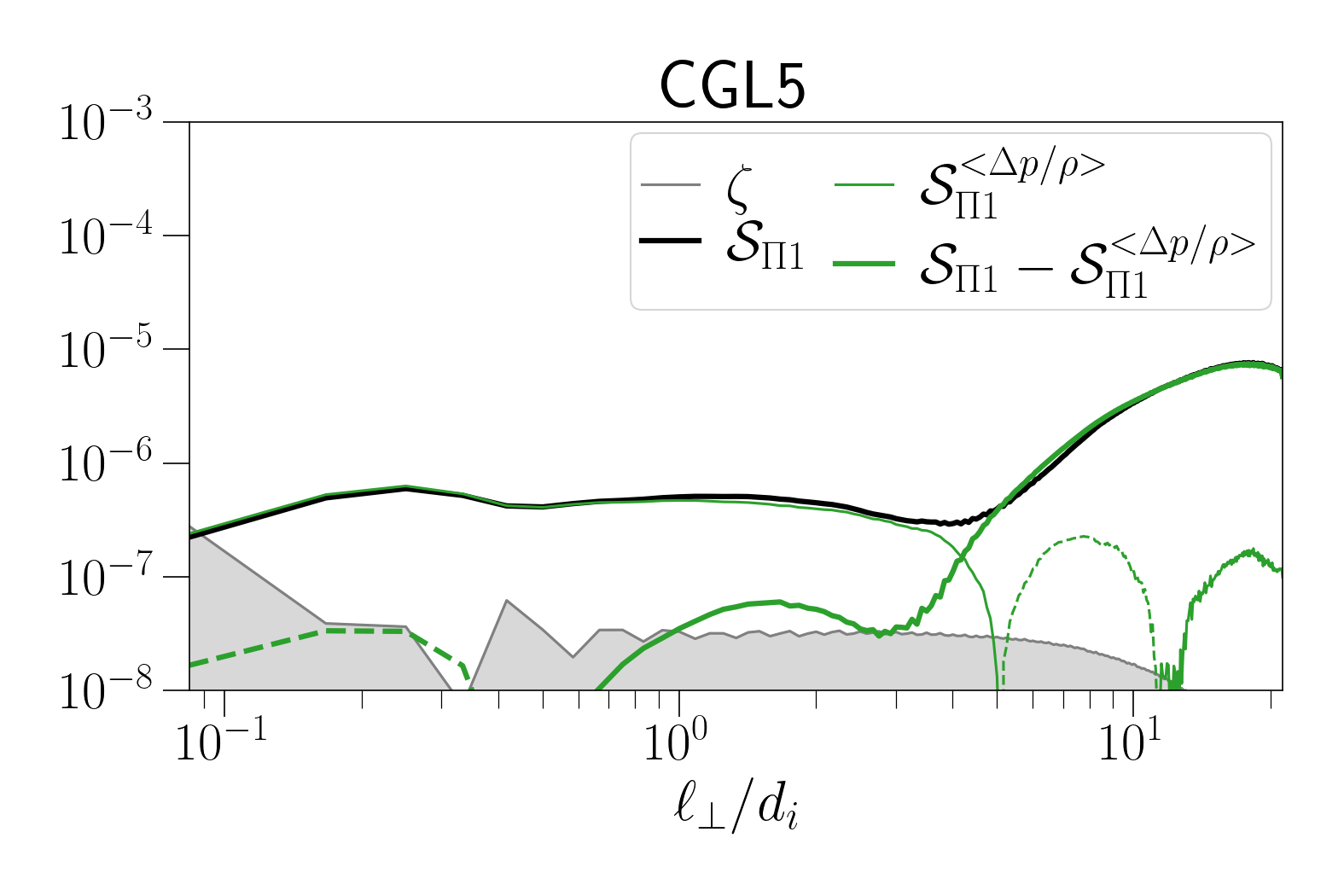}
\includegraphics[width=0.49\textwidth,trim = 1cm 1cm 1cm 1cm, clip]{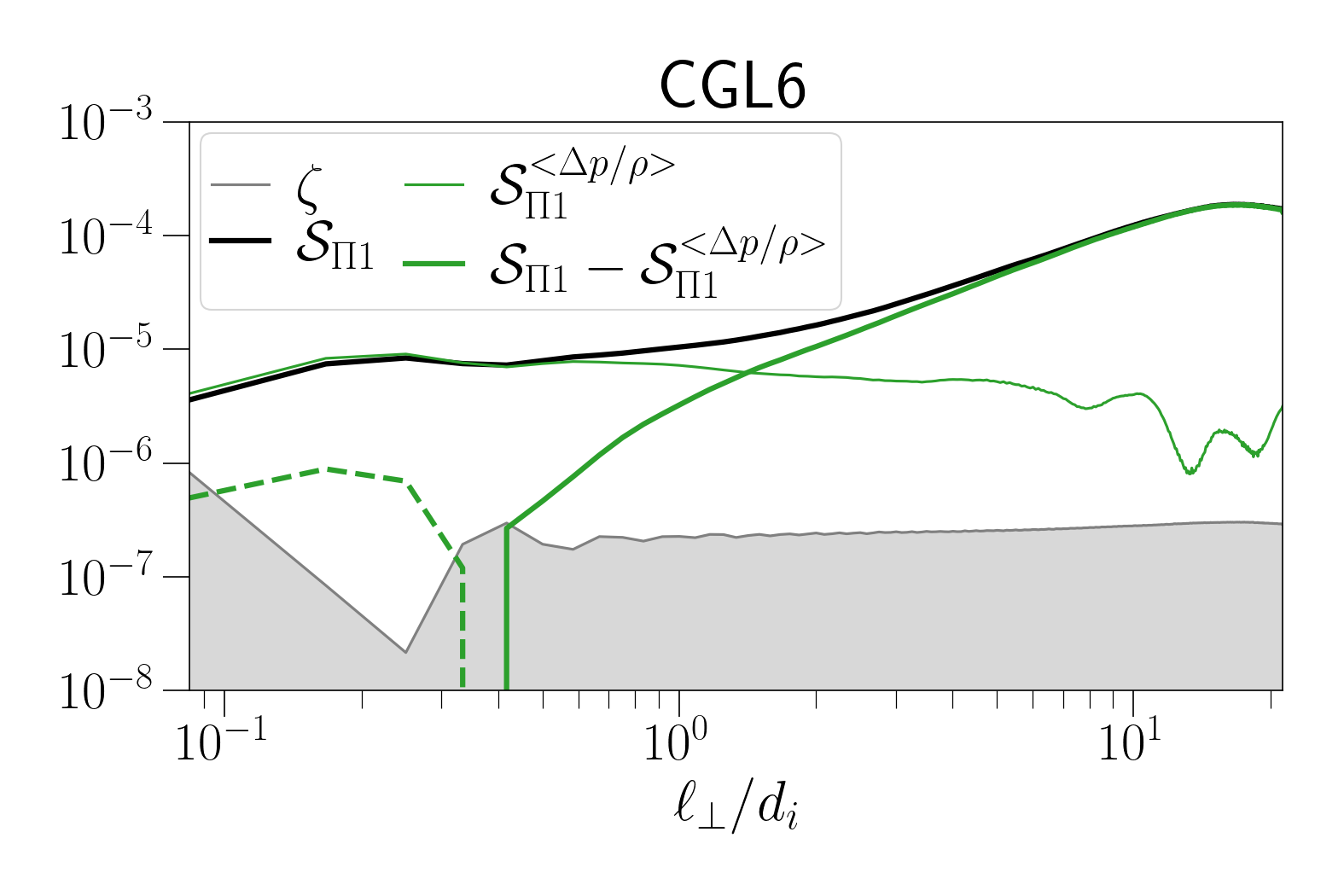}
\caption{For each simulation, 1D perpendicular profile of the decomposition of $\mathcal{S}_{\Pi 1}$ into terms depending or not on the (mean) quantity $\left< \frac{\Delta p}{ \rho} \right> = \left< \frac{p_{\parallel}-p_{\perp}}{ \rho} \right>$.}
\label{fig:spi1_dec}
\end{figure}

To compare the impact of the averaged pressure anisotropy to the one of its fluctuations, we decompose $\mathcal{S}_{\Pi 1}$ into $\mathcal{S}_{\Pi 1}^{\langle \frac{\Delta p}{\rho}\rangle} = 2 \langle \frac{\Delta p}{\rho} \rangle \langle \delta \left(\boldsymbol{bb}\right): \left(\rho'\bnabla \boldsymbol{v} - \rho \bnabla' \boldsymbol{v'}\right) \rangle$ with $\Delta p = p_{\parallel} - p_{\perp}$ and $\mathcal{S}_{\Pi 1} - \mathcal{S}_{\Pi 1}^{\langle \frac{\Delta p}{\rho}\rangle}$. The results are displayed in Fig. \ref{fig:spi1_dec}. For all simulations, the quantity depending on the fluctuations of $\Delta p=p_\parallel-p_\perp$ dominates the inertial range, while that depending on the mean value dominates the scales closest to the forcing. For CGL1, CGL2, and CGL3B, the change of sign is associated to a sign variation of the contribution depending on the fluctuations of $\Delta p$. Note that, similarly to the cascade rate \eqref{eq:extL}, the pressure always appears divided by the density, e.g., $\frac{\Delta p}{\rho}$. Thus, the observed behavior could reflect the temperature rather than the pressure anisotropy, although they almost coincide in the present simulations, due to the weak density fluctuations.

\section{Discussion}\label{sec:disc}

\subsection{Spectra of the pressure anisotropy} 

\begin{figure}
 \centering
 \includegraphics[width=0.6\textwidth]{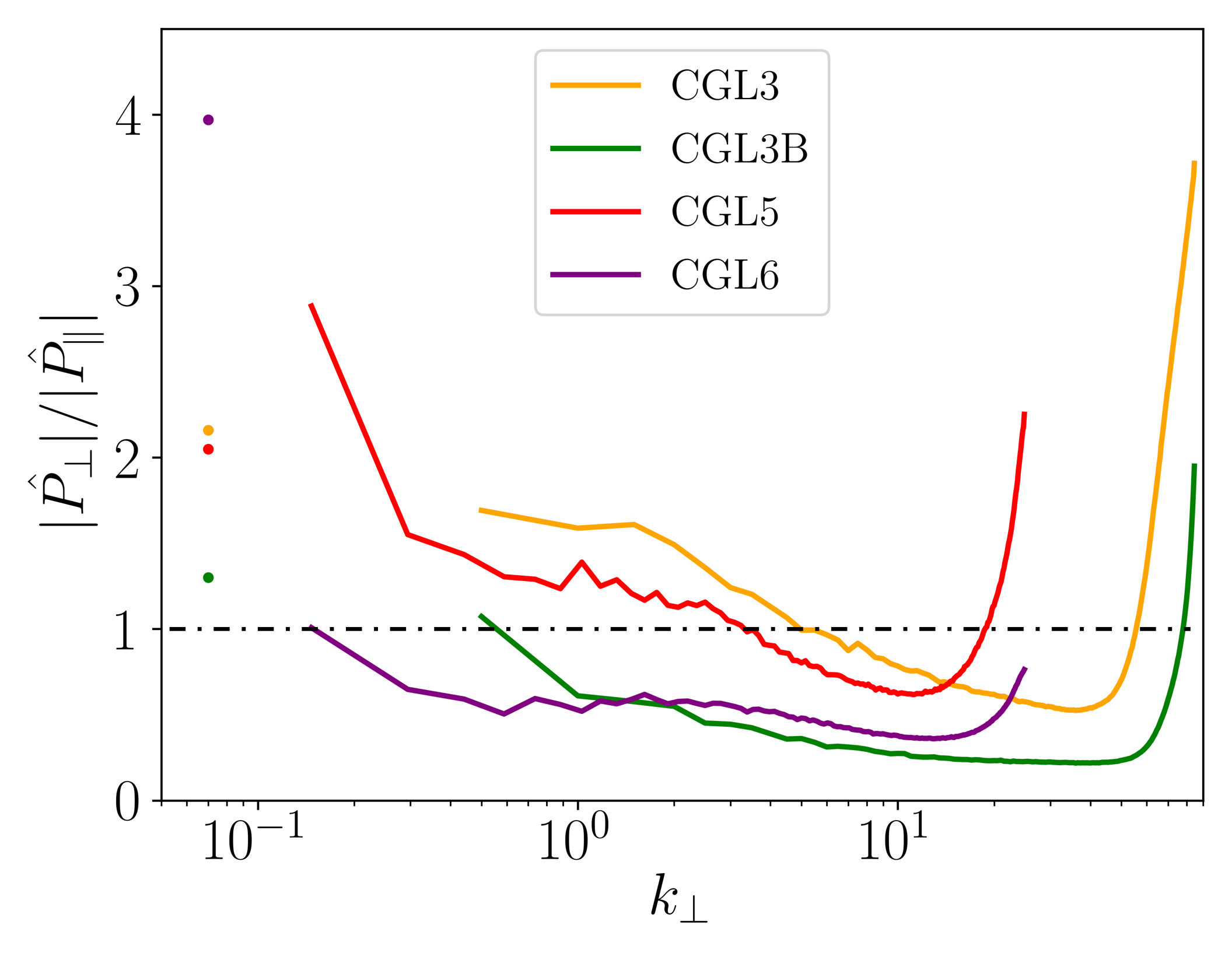}
 \caption{Ratio of the spectral distributions of the perpendicular and parallel pressures in the transverse Fourier plane, 
 $|{\hat P}_\perp|/|{\hat P}_\| |$, where ${\displaystyle |{\hat P}_\perp| (k_{\perp}) =\sqrt{\sum_{k_\|} \sum_{k_\perp\le |{\boldsymbol k}'_\perp |\le k_\perp + \Delta k_\perp} |{\widehat p}_\perp({\boldsymbol k}'_\perp, k_\|)|^2}}$ and ${\displaystyle |{\hat P}_\| |(k_{\perp}) =\sqrt{\sum_{k_\|} \sum_{k_\perp\le |{\boldsymbol k}'_\perp| \le k_\perp + \Delta k_\perp} |{\widehat p}_\|({\boldsymbol k}'_\perp, k_\|)|^2}}$, plotted as a function of $k_\perp$, for runs CGL3 (yellow), CGL3B (green), CGL5 (red) and CGL6 (purple). Here, ${\widehat p}_\perp$ and ${\widehat p}_\|$ hold for the Fourier transform of the perpendicular and parallel pressures $p_\perp$ and $p_\|$, respectively. For comparison, the colored points on the left part of the figure indicate the ratio of the mean perpendicular and parallel pressures for the corresponding simulations.}
 \label{fig:spectra_p}
\end{figure}

Section \ref{sec:results} shows that the pressure-anisotropy contribution to the total cascade rate in the inertial range is led by the fluctuations of the pressure anisotropy. To better interpret this observation, a measure of the ratio of the spectral distributions of the perpendicular and parallel pressures for runs CGL3, CGL3B, CGL5, and CGL6 is presented in Fig. \ref{fig:spectra_p}. These runs have been chosen because they show the largest contribution of the pressure-anisotropy terms to the cascade rate (CGL3B and CGL6 in the inertial range; CGL3, CGL5, and CGL6 at the largest scales).

We observe a change of pressure anisotropy, from $|{\hat P}_\perp|(k_\perp)>|{\hat P}_\|| (k_\perp)$ (see definitions in the figure caption) to the opposite ordering, for runs CGL3 and CGL5 at $k_\perp d_i\sim 4$. Differently, CGL3B and CGL6 show a monotonic behavior ($|{\hat P}_\perp|(k_\perp)<|{\hat P}_\|| (k_\perp)$) at nearly all scales (excluding the dissipation range), with the strongest pressure anisotropy at the smallest scales. We acknowledge the difficulty of relating the behavior of this quantity to the complex nonlinear terms of the third-order law that involve pressure anisotropy. Nevertheless, we can speculate that the strong and uniform-in-scale pressure anisotropy observed for runs CGL3B and CGL6 could explain (or at least, is consistent with) the enhanced contribution of $\varepsilon_{\overline{\boldsymbol{\Pi}}}$ observed in the inertial range for those runs in comparison to the others. On the other hand, the change from a perpendicular-pressure dominated regime to a parallel-pressure dominated one observed for runs CGL3 and CGL5, with potential fluctuations in sign of $\Delta p$, may result in contributions that would statistically cancel out, hence the weak contribution of the anisotropic terms in the cascade rate for those runs. 

\subsection{Plasma-state distribution in the $(a_p, \beta_\|)$-plane}

\begin{figure}
\center
\includegraphics[width=0.8\textwidth]{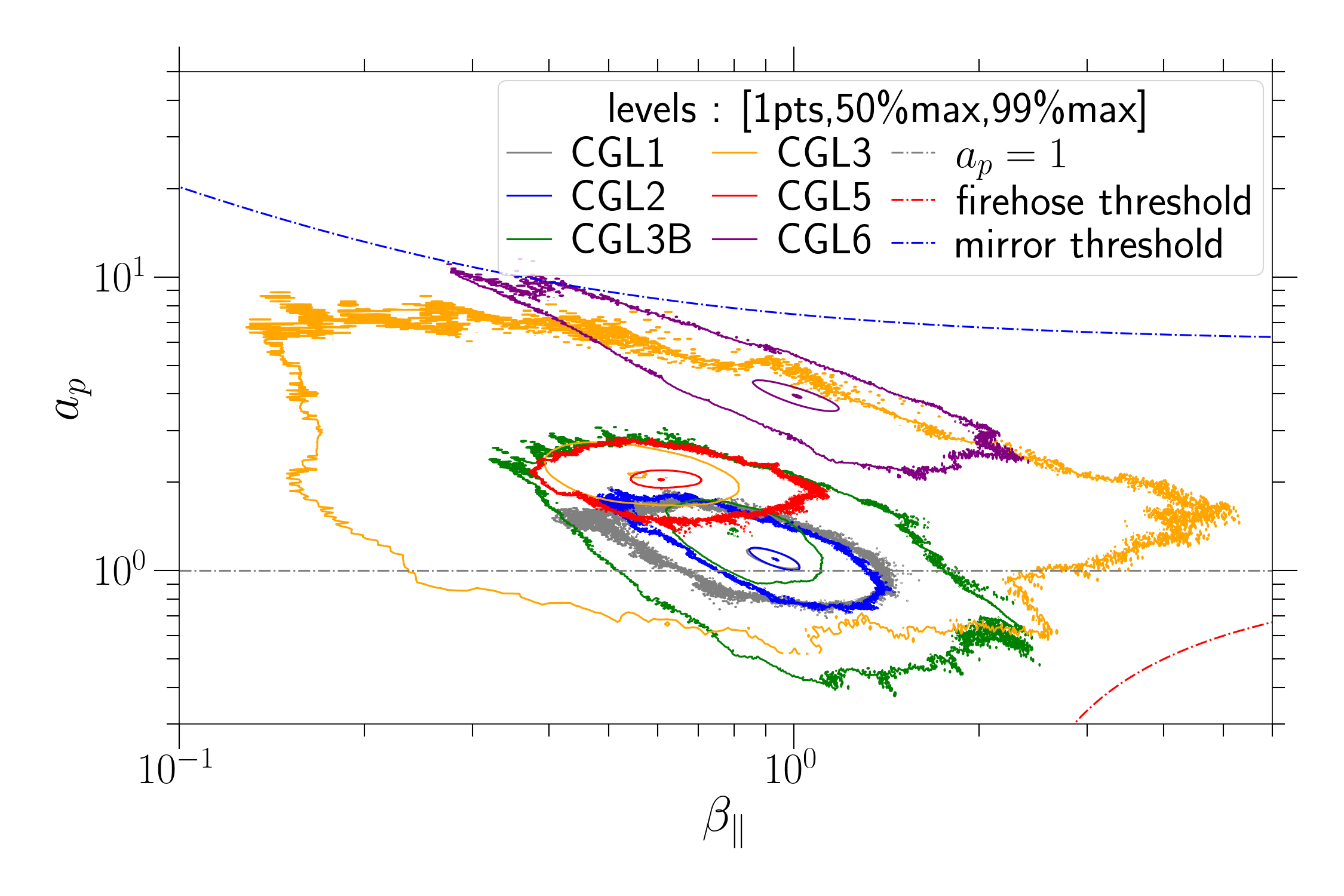}

\caption{Distribution of the plasma states in the $(a_p,\beta_{\parallel})$-plane with their occurrence rate: occurrence points (edge of the distribution), $50\%$ (inner ellipse) and $99\%$ (outer curves points) in the various simulations. Dash-dotted lines indicate the firehose (orange) and mirror (blue) linear instability thresholds (taking into account the isothermal electronic pressure).}
\label{fig:brazil}
\end{figure}
\begin{table}
\begin{center}
\begin{tabular}{ c|cccccc } 
Name & CGL1 & CGL2 & CGL3 & CGL3B & CGL5 & CGL6 \\
\hline
$\left<\rho\right> \pm \sigma_{\rho}$ & $1.00 \pm 0.02$ & $1.00 \pm 0.02$ & $1.00 \pm 0.08$ & $1.00 \pm 0.04$ & $1.00 \pm 0.02$ & $1.00 \pm 0.03$ \\
\hline
$\left<\beta_{i} \right> \pm \sigma_{\beta_i}$ & $0.99 \pm 0.07$ & $0.99 \pm 0.06$ & $1.13 \pm 0.39$ & $1.03 \pm 0.19$ & $1.07 \pm 0.12$ & $2.98 \pm 0.25$\\
$\left<\beta_{\parallel } \right> \pm \sigma_{\beta_{\parallel}}$ & $0.93 \pm 0.09$ & $0.93 \pm 0.08$ & $0.65 \pm 0.25$ & $0.87 \pm 0.18$ & $0.63 \pm 0.07$ & $1.01 \pm 0.15$ \\
\hline
$\left<a_{p} \right>\pm \sigma_{a_p}$ & $1.11 \pm 0.08$ & $1.11 \pm 0.10$ & $2.16 \pm 0.49$ & $1.30 \pm 0.30$ & $2.05 \pm 0.12$ & $3.97 \pm 0.47$\\
$a_p>1$(in \%) & 91 & 88 & 99.8 & 83 & 100 & 100\\
$a_p<1$(in \%) & 9 & 12 & 0.2 & 17 & 0 & 0 \\
\hline
$\left<- \beta_{\parallel }^* \right>\pm \sigma_{\beta_{\parallel}^*}$ & $0.09 \pm 0.07$ & $0.09 \pm 0.08$ & $0.71 \pm 0.32$ & $0.24 \pm 0.24$ & $0.66 \pm 0.10$ & $2.96 \pm 0.25$ \\
\end{tabular}
\caption{Average and standard deviation of density $\rho$ (or electronic pressure), total ionic beta-parameter $\beta_i = (2p_{\perp}+p_{\parallel})/(3p_m)$ and parallel beta-parameter $\beta_{\parallel} = p_{\parallel}/p_m$, pressure anisotropy $a_p$, percentage of data points where $a_p>1$ or $a_p<1$, and parameter $-\beta_{\parallel}^* = \beta_{\parallel}(a_p-1)$, entering the terms of $\varepsilon_{\overline{\boldsymbol{\Pi}}}$ (see \eqref{eq:dtail_an})}.
\label{tab:stat_anis}
\end{center}
\end{table}

The distributions of the plasma representative points in the $(a_p,\beta_{\parallel})$ diagram is displayed in Fig. \ref{fig:brazil}, along with the thresholds of the linear firehose and mirror instabilities (see Appendix \ref{ann:insta}). For all the runs, we observe that the representative points are localized in the stable region, except for a few points from CGL6 laying near the mirror-instability threshold. One can notice that the distributions associated with CGL1 to CGL5 are not centered around the initial values $a_{p0} =1$ and $\beta_{\parallel 0} = 1$, but are shifted to a new regime of parameters. The representative points from simulations CGL1, CGL2, and CGL3B stay close to $a_{p0}=1$, while those corresponding to CGL3 and CGL5 shift up to an average value $\left<a_{p}\right> = 2$, with, however, a larger standard deviation for the former run as shown in Table \ref{tab:stat_anis}. This migration is probably due to a direct conversion at large scales of part of the injected energy into internal energy, thus enhancing the average temperatures without contributing to $\varepsilon_{iso}$, as shown by \citet{ferrand_fluid_2021} with CGL1, CGL2 and CGL3 simulations. The difference in the migration appears to be consistent with the results reported above about the importance of the contribution of $\varepsilon_{\overline{\boldsymbol{\Pi}}}$ to the total cascade rate $\varepsilon_{cgl}$ at large scales where the contribution of $\lan\frac{\Delta p}{\rho}\ran$ is dominant. However, the proportion of points with $a_p<1$, higher for CGL1, CGL2 and CGL3B, could explain the change of sign of $\varepsilon_{\overline{\boldsymbol{\Pi}}}$, as relatively more points with $\frac{\Delta p}{\rho} < 0$ contribute to $\mathcal{S}_{\Pi 1} - \mathcal{S}_{\Pi 1}^{\langle \frac{\Delta p}{\rho}\rangle}$. More detailed statistical studies using spacecraft data from the solar wind, for instance, will be necessary to disentangle the behavior of the cascade rate and the localization of the plasma state in the $(a_p,\beta_{\parallel})$-plane and to validate the correlations reported here.

\section{Conclusion}

This paper summarizes the results of a study of the impact of pressure anisotropy on the turbulent cascade rate described by the third-order law obtained in \citet{simon_exact_2022}, using six simulations of CGL-Hall-MHD turbulence. The contribution to the cascade rate of the terms independent of the anisotropic component of the pressure tensor has been found to agree with the results of previous studies using incompressible or weakly compressible models with isotropic pressure. In addition, the contribution of the anisotropic component of the pressure tensor reveals
\begin{itemize}
 \item the relevance of this effect even in a quasi-incompressible case, 
 \item an impact of the fluctuations of pressure anisotropy (visible both in real and Fourier space) on the cascade rate in the inertial range,
 \item a more important effect close to the forcing scales, linked to the anisotropy of the mean pressure, 
 \item a sign variation of the contribution of $\varepsilon_{\overline{\boldsymbol{\Pi}}}$ to the cascade rate, potentially linked to the localization of the simulation in the $(\beta_{\parallel},a_p)$ diagram. 
\end{itemize}
 Even if CGL6, which is initialized with ${a_p}_0=4$, shows a more prominent contribution of the pressure anisotropy term, the impact in the inertial range remains moderate. Extension of this study would require other simulations with different ${a_p}$ and possibly higher compressibility. Also, the third-order law of \citet{simon_exact_2022} is derived with a particular form of correlation function $\mathcal{R}$. This choice is not unique, and alternative quantities have been considered in the literature \cite{ferrand_compressible_2020,hellinger_ion-scale_2022}. The sensitivity of the present conclusions to the retained correlation function is to be analyzed in further studies.

On the other hand, spacecraft data can also be used to conduct further studies, as they give access to a broad range of parameter space $(\beta_\parallel,a_p,\delta \rho/\rho)$. However, computing pointwise (spatial) derivatives involved in the third-order law using multi-spacecraft data (e.g., Cluster or MMS) requires some caution as the uncertainties can be too large to allow reliable quantification of the small contributions due to density fluctuations or pressure anisotropy. The nine probes of the upcoming NASA/Helioswarm mission is likely to permit improving those estimates. 

\section{Contribution and acknowledgement}

P.S. is funded by a DIM-ACAV + Doctoral fellowship
(2020-2023). She performed the present study. FS and SG supervised the work. The numerical simulation setups were designed and data were obtained in collaboration with D. Laveder, T. Passot, and P.L. Sulem who also provided Figs. \ref{fig:3D}, \ref{fig:Bperp_spectra} and \ref{fig:spectra_p}. All authors contributed to discussing the results and writing the paper. PS thanks A. Jeandet for his help in the implementation of the data post-processing. Computations were performed at the mesocenter Cholesky of Ecole Polytechnique and mesocentre SIGAMM, hosted by Observatoire de la Côte d'Azur (OCA). 

\appendix

\section{CGL firehose and mirror instability thresholds in the presence of isothermal electrons} \label{ann:insta}

To obtain the CGL-MHD-instability thresholds with isothermal electrons for the firehose and mirror instabilities, we linearize the dimensionless system \eqref{eq:model}, taking $d_i=0$ (no Hall effect) and neglecting dissipative and forcing terms. The mean component of each quantity is isolated and indexed by the subscript $0$. Furthermore, at equilibrium, $\mathbf{v}_0 = 0$ and $\mathbf{B_{0}} = [0,0,B_{0}]_{x,y,z}$ are assumed. The electronic pressure is supposed isothermal ($\propto \rho$), consistent with the numerical simulations. Then, we apply the transformation to the Fourier space ($\partial_t \rightarrow -i\omega$, $\partial_x \rightarrow i k_{\perp}$, $\partial_y \rightarrow i k_y$, $\partial_z \rightarrow i k_{\parallel}$) with $k_y = 0$. The velocity equations become
 \begin{eqnarray}
 \left\{
 \begin{split}
 0 &= - \frac{1}{v^2_{A0}}\omega v_x + \frac{\beta_{\parallel}}{2} a_p k_{\perp} \frac{p_{\perp}}{p_{\perp 0}} + (\frac{\beta_{\parallel}}{2} (1 - a_p) -1) k_{\parallel} \frac{B_x}{B_0} +k_{\perp} \frac{B_z}{B_0} +\frac{\beta_{e}}{2} k_{\perp} \frac{p_e}{p_{e0}} \\
 0 &=- \frac{1}{v^2_{A0}} \omega v_y + (\frac{\beta_{\parallel}}{2} (1 - a_p) -1)k_{\parallel} \frac{B_y}{B_0} \\
 0 &= - \frac{1}{v^2_{A0}} \omega v_z +\frac{\beta_{\parallel}}{2} k_{\parallel} \frac{p_{\parallel}}{p_{\parallel 0}} + \frac{\beta_{\parallel}}{2} (1 - a_p) k_{\perp} \frac{B_x}{B_0} + \frac{\beta_{e}}{2} k_{\parallel} \frac{p_e}{p_{e0}}
\end{split} 
 \right. \label{eq:linv}
 \end{eqnarray}
 with $v^2_{A0} = \frac{B^2_0}{4\pi \rho_0}$, $\beta_{\parallel} = \beta_0 \frac{4\pi p_{\parallel 0}}{B^2_0} $ and $a_p = \frac{p_{\perp 0}}{p_{\parallel 0}}$, $D_i = \frac{d_i}{\rho_0}$, $\beta_{e} = \beta_0 \frac{4\pi p_{e 0}}{B^2_0} $.

The other equations read
\begin{eqnarray}
\left\{
 \begin{split}
 0 &= -\omega \rho + \rho_0 ( k_{\perp} v_x + k_{\parallel} v_z) \\
 0 &= -\omega B_x - B_0 k_{\parallel} v_x \\ 
 0 &= -\omega B_y - B_0 k_{\parallel} v_y \\
 0 &= -\omega B_z + B_0 k_{\perp} v_x 
\end{split} 
\right.\nonumber
\end{eqnarray}
\begin{eqnarray}
\left\{
 \begin{split}
 0 &= -\omega p_{\parallel} + p_{\parallel 0} (k_{\perp} v_x + 3 k_{\parallel} v_z) \\
0 &= -\omega p_{\perp} + p_{\perp 0} (2k_{\perp} v_x +k_{\parallel} v_z) \\
 0 & = -\omega p_{e} + p_{e0} ( k_{\perp} v_x + k_{\parallel} v_z) 
\end{split} 
\right. \nonumber \\ \label{eq:linq}
\end{eqnarray}

Injecting \eqref{eq:linq} in \eqref{eq:linv} gives
\begin{eqnarray}
\left\{
 \begin{split}
0 &= ( - \frac{1 }{v^2_{A0}}\omega^2+(2\frac{\beta_{\parallel}}{2} a_p +1+ \frac{\beta_e}{2})k^2_{\perp} - (\frac{\beta_{\parallel}}{2} (1 - a_p) -1) k^2_{\parallel} )v_x + (\frac{\beta_{\parallel}}{2} a_p+ \frac{\beta_e}{2}) k_{\perp}k_{\parallel}v_z \\
 0 &= (- \frac{1}{v^2_{A0}} \omega^2 - (\frac{\beta_{\parallel}}{2} (1 - a_p) -1)k^2_{\parallel} ) v_y \\
 0 &=(\frac{\beta_{\parallel}}{2} a_p+ \frac{\beta_e}{2}) k_{\perp} k_{\parallel} v_x + (- \frac{1}{v^2_{A0}} \omega^2 + (3\frac{\beta_{\parallel}}{2} + \frac{\beta_e}{2}) k^2_{\parallel} ) v_z 
 \end{split} 
\right. \label{eq:linv_v2}
\end{eqnarray}

The determinant of the system \eqref{eq:linv_v2} involves two factors. The first one gives the dispersive relation of the incompressible Alfv\'en-wave, polarized such as $v_x = v_z = 0$ i.e. $\omega = \pm k_{\parallel} v_{A0} \sqrt{1-(\frac{\beta_{\parallel}}{2} (1 - a_p) )} $. The Alfvén wave is affected by the firehose instability that appears if \[1-(\frac{\beta_{\parallel}}{2} (1 - a_p) ) < 0.\]

The second factor gives the dispersive relation of the magnetosonic waves polarized such as $v_y = 0$, and takes the form of a second order polynomial $\omega^4 + A_1 \omega^2 + A_0 =0 $ with 
\[A_1 = - v^2_{A0} ((2\frac{\beta_{\parallel}}{2} a_p +1+ \frac{\beta_e}{2})k^2_{\perp} +(\frac{\beta_{\parallel}}{2} (2+a_p) +1+ \frac{\beta_e}{2}) k^2_{\parallel} ) \]
\[A_0 = v^2_{A0} k^2_{\parallel} ((\frac{\beta_{\parallel}}{2}(\frac{\beta_{\parallel}}{2}a_p (6 - a_p) +3)+ \frac{\beta_e}{2} (3\frac{\beta_{\parallel}}{2} +1) ) k^2_{\perp} - (3\frac{\beta_{\parallel}}{2} + \frac{\beta_e}{2})(\frac{\beta_{\parallel}}{2} (1 - a_p) -1) k^2_{\parallel}) .\]
Assuming that $k_{\perp}\gg k_{\parallel}$, $A_0$ gives the mirror threshold altered by the isothermal electronic pressure: 
\[\frac{\beta_{\parallel}}{2}a_p (6 - a_p) +3 + \frac{\beta_e}{\beta_{\parallel}} (3\frac{\beta_{\parallel}}{2} +1) = 0\]
If $\beta_e =0$, the threshold for the CGL-MHD \citep{hunana_introductory_2019} is recovered. In our simulations, $\beta_e = 1$.


%

\end{document}